\pdfoutput=1
  \def\draftversion{false}
  
\documentclass[aps,prb,preprint,showpacs,amsmath,amssymb,superscriptaddress,altaffilletter,longbibliography,floatfix]{revtex4-2}

\usepackage{footnote} 
\usepackage{tabularx}
\usepackage{tikz}
\usepackage{float}
\usepackage{graphicx}
\usepackage{lipsum} 
\usepackage{wrapfig}
\usepackage{ifthen}
\usepackage{color} 
\usepackage{bm} 
\usepackage{multirow}
\usepackage{hyperref} 
\usepackage{bbold} 

\ifthenelse{\equal{\draftversion}{true}}{
  \marginparwidth 2.7in
  \marginparsep 0.5in
  \newcounter{comm} 
  \def\commnext{\stepcounter{comm}}
  \def\commtext{{\bf\color{blue}[\arabic{comm}]}}
  \def\commmar{{\bf\color{blue}[\arabic{comm}]}}
  \def\jlm#1{\commnext\marginpar{\small JL\commmar: #1}\commtext}
  \def\jdm#1{\commnext\marginpar{\small JRD\commmar: #1}\commtext}
  \def\bbm#1{\commnext\marginpar{\small BAB\commmar: #1}\commtext}
}{
  \def\jlm#1{}
  \def\jdm#1{}
  \def\bbm#1{}
}

\def\nn{\nonumber\\}
\def\vr{\mathbf{r}}

\def\vq{\mathbf{q}}
\def\vv{\mathbf{v}}
\def\kp{\mathbf{k}}
\def\Gp{\mathbf{G}}

\def\fo{\hat{\psi}}
\def\fod{\hat{\psi}^{\dagger}}
\def\SOC{\textrm{SOC}}
\def\CF{\textrm{CF}}
\def\KS{\textrm{KS}}
\def\xc{\textrm{xc}}
\def\Sop{\hat{S}}
\def\nn{\nonumber\\}
\def\d{\textrm{d}}
\def\ii{\textrm{i}}

\def\vv{\mathbf{v}}
\def\vr{\mathbf{r}}

\def\vk{\mathbf{k}}

\def\vq{\mathbf{q}}
\def\btau{\boldsymbol{\tau}}
\def\kp{\mathbf{k}}

\def\rp{\mathbf{r}}
\def\Gp{\mathbf{G}}
\def\rot{\mathrm{R}}

\def\nx{n_{\textrm{x}}}
\def\ny{n_{\textrm{y}}}
\def\nz{n_{\textrm{z}}}
\def\bsp{\begin{split}}
\def\esp{\end{split}}
\def\nn{\nonumber\\}
\def\btau{\boldsymbol{\tau}}

\def\rot{\mathrm{R}}

\def\SOC{\textrm{SOC}}
\def\CF{\textrm{CF}}
\def\KS{\textrm{KS}}

\def\xc{\textrm{xc}}

\def\KS{\textrm{KS}}
\def\COH{\textrm{COH}}
\def\SX{\textrm{SX}}

\def\Bi2Se3{\textrm{Bi}_2\textrm{Se}_3}

\begin{document}

\title{Fully relativistic $GW$/Bethe-Salpeter calculations in BerkeleyGW: implementation, symmetries, benchmarking, and performance}

\author{Bradford A. Barker}
\email{bbarker6@ucmerced.edu}
\affiliation{Department of Physics, University of California,
  Berkeley, Berkeley, CA 94720, USA}
\affiliation{Materials Sciences Division, Lawrence Berkeley National
  Laboratory, Berkeley, CA 94720, USA}
\affiliation{Department of Physics, University of California,
  Merced, Merced, CA 95343, USA}
\author{Jack Deslippe}
\affiliation{National Energy Research Scientific Computing Center, Lawrence Berkeley National Laboratory, Berkeley, CA 94720, USA}
\author{Johannes Lischner}
\affiliation{Department of Materials and Physics, and Thomas Young Centre for Theory and Simulation of Materials,
Imperial College London, London SW7 2AZ, United Kingdom}
\author{Manish Jain}
\affiliation{Center for Condensed Matter Theory, Department of Physics, Indian Institute of Science, Bangalore 560012, India}
\author{Oleg V. Yazyev}
\affiliation{Institute of Physics, Ecole Polytechnique Federale de Lausanne (EPFL), CH-1015 Lausanne, Switzerland }
\author{David A. Strubbe}
\affiliation{Department of Physics, University of California, Merced,
  Merced, CA  95343, USA}
\author{Steven G. Louie} 
\affiliation{Department of Physics, University of California,
  Berkeley, Berkeley, CA 94720, USA}
\affiliation{Materials Sciences Division, Lawrence Berkeley National
  Laboratory, Berkeley, CA 94720, USA}

\date{\today}


\begin{abstract}
Computing the $GW$ quasiparticle bandstructure and Bethe-Salpeter Equation (BSE) absorption spectra for materials with spin-orbit coupling
has commonly been done by treating $GW$ corrections and spin-orbit coupling as separate perturbations to density-functional theory.
However, accurate treatment of materials with strong spin-orbit coupling (such as many topological materials of recent interest, and thermoelectrics)
often requires a fully relativistic approach using spinor wavefunctions in the Kohn-Sham equation and $GW$/BSE.
Such calculations have only recently become available, in particular for the BSE.
We have implemented this approach in the plane-wave pseudopotential $GW$/BSE code BerkeleyGW, which is highly parallelized and
widely used in the electronic-structure community.
We present reference results for quasiparticle bandstructures and optical absorption spectra of
solids with different strengths of spin-orbit coupling, including
Si, Ge, GaAs, GaSb, CdSe, Au, and Bi$_2$Se$_3$.
The calculated quasiparticle band gaps of these systems are found to agree
with experiment to within a few tens of meV. SOC splittings are found to be generally in better agreement with experiment, including quasiparticle corrections to band energies.
The absorption spectrum of GaAs is not significantly impacted by the inclusion of spin-orbit coupling due to its relatively small value (0.2~{eV}) in the $\Lambda$ direction, while the absorption spectrum
of GaSb calculated with the fully-relativistic $GW$-BSE 
captures the large spin-orbit splitting of peaks in the spectrum.
For the prototypical topological insulator Bi$_2$Se$_3$,
we find a drastic change in the low-energy bandstructure compared to that of DFT, with the fully-relativistic
treatment of the $GW$ approximation correctly capturing the parabolic nature of
the valence and conduction bands after including off-diagonal self-energy matrix elements.
We present the detailed methodology, approach to spatial symmetries for spinors, comparison against other codes, and performance
compared to spinless $GW$/BSE calculations and perturbative approaches to SOC. This work aims to spur further development of spinor $GW$/BSE methodology in excited-state research software, and enables more accurate and
detailed exploration of electronic and optical properties of materials containing elements with large atomic number.
\end{abstract}

\maketitle

\section{Introduction}

Solid state physics and materials research is increasingly focusing its attention
on materials containing heavy elements. Such materials have large spin-orbit coupling,
often exceeding 1~{eV} for atoms from the fifth and sixth rows of the periodic table.
These materials are important as thermoelectrics\cite{ceder,zhao2014bicuseo,pei2012band,sootsman2009new,chen2003recent,zhao2014review,chung2004new,wood1988materials} and also can be topological insulators 
\cite{haldane,kane_mele_qshe_graphene,hasan,moore2010birth,fu,konig,bernevig,murakami,ando}
and Weyl semi-metals
\cite{wan2011semimetal,burkov2011weyl,lv2015experimental,weng2015weyl},
  among other novel topological phases\cite{liu2014na3bi,schindler2018higher,tanaka2012tci}.
Hybrid organic metal halide perovskite materials are also of great interest for photovoltaics,
and contain heavy elements such as Pb, I\cite{solvent_mapbi3}, and/or Bi\cite{bismuth_perovskite},
and spin-orbit effects like Rashba splitting can play a role in their optical properties \cite{frohna}.
%
The standard approach to investigating the ground state electronic structure of these materials
is Density Functional Theory (DFT)\cite{kohn_sham,hohenberg_kohn}.
Despite its widespread use to compute bandstructures, it is important to note
that the Kohn-Sham eigenvalues of DFT do not have a rigorous physical meaning apart from the energy of the highest occupied molecular orbital,
resulting in the well-known band gap problem of DFT.
To compute excited-state properties such as bandstructures
and absorption spectra, one must go beyond DFT and use many-body perturbation theory
approaches, such as the $GW$\cite{hedin,hybertsen} and $GW$-BSE methods\cite{rohlfing}.



For materials with weak spin-orbit coupling, quasiparticle bandstructures incorporating spin-orbit coupling
can be computed by separately calculating the additional contribution to the energy eigenvalues from spin-orbit coupling via conventional perturbation theory, in which the Hamiltonian
is constructed using bands $|n\kp\rangle_0$ that have been computed previously while neglecting spin-orbit coupling\cite{hybertsen_soc,hemstreet}, by diagonalizing the Hamiltonian
\begin{equation}
    \label{eq:gw_plus_soc}
    H_{n_1,\kp,\alpha;n_2,\kp,\beta} = \langle n_1 \kp |_0 \, \langle \alpha |\,  E_{n_1\kp,\, 0}^{\textrm{QP}} \, \delta_{n_1 n_2}\delta_{\alpha \beta} + H^{\SOC}_{\alpha \beta} \, | \beta \rangle \, | n_2 \kp \rangle_0 \, ,
\end{equation}
with the subscript ``0'' denoting quantities that neglect spin, and $|\alpha \rangle$ and $|\beta \rangle $ are spinor basis states, $|\uparrow\rangle = \begin{pmatrix}1 , & 0\end{pmatrix}^T$ or $|\downarrow\rangle = \begin{pmatrix}0 , & 1\end{pmatrix}^T$.
This approach,
``$GW$+SOC,'' has been successfully used
in $ab$ $initio$ calculations of diamond- and zinc-blende-structure semiconductors\cite{malone}, metals such as Au\cite{tonatiuh}, and topological insulators $\Bi2Se3$ \cite{yazyev} and Bi$_2$Te$_3$ \cite{kioupakis},
among other systems. When the Kohn-Sham bandstructure neglecting spin-orbit coupling is qualitatively similar to the quasiparticle bandstructure that includes it, the $GW$+SOC approach is generally sufficient. 
Despite the success of perturbation theory in computing the changes of eigenvalues
for materials with weak spin-orbit coupling, there is a clear need for a non-perturbative first-principles
treatment of materials with strong spin-orbit coupling.
In particular, some materials containing heavy elements, such as Bi$_2$Se$_3$\cite{zhang}
and $\beta$-HgS\cite{hgx_soc,my_hgs}, have DFT bandstructures that change
significantly when spin-orbit coupling is included.
In cases such as these, the perturbative $GW$+SOC approach is quantitatively or even qualitatively inaccurate.
One should use a fully-relativistic treatment from the outset,
starting with the calculation of the two-component spinor Kohn-Sham states and then using these states
to calculate excited-state properties, such as the quasiparticle bandstructure and the absorption spectrum.
This first-principles method also allows for capturing the effect of the renormalization
of the spin-orbit coupling strength \cite{sakuma}, along with improved band gaps.

Due to the doubled number of bands and doubled size of the wavefunctions compared to spinless calculations, there is a significant increase
in the already substantial computational expense of many-body perturbation theory calculations, not to mention a significant increase in the complexity of the computer code. As a result, fully relativistic spinor calculations
with $GW$ have only recently become available, and used in the literature.
The all-electron FLAPW code SPEX's implementation \cite{spex} was later followed by pseudopotential and PAW codes (WEST\cite{west}, Yambo\cite{yambo}, 
FHIaims \cite{fhiaims}, GPAW \cite{gpaw_soc}, and VASP\cite{vasp}).
There are yet fewer spinor BSE codes available; to date, only Yambo \cite{yambo_bse_soc} and BerkeleyGW have the capability to solve the Bethe-Salpeter Equation with two-component spinor wavefunctions. While plane-wave DFT codes have become highly
comparable in recent years due to increasing consensus on the best algorithms to use, and great efforts to determine
the source of any discrepancies \cite{Lejaeghere}, there is a significant variation in the approaches used in
$GW$/BSE codes, including not only basis sets and pseudopotentials, but also plasmon pole models, frequency integration,
interpolation schemes, handling of the dielectric matrix, acceleration of sums over empty states, solution of Dyson's equation,
and other numerical tricks and details. Such details are only
sometimes spelled out comprehensively for a given code \cite{BerkeleyGW}. Benchmarking projects for $GW$ codes
-- and especially for BSE -- are still in their infancy. A notable example is the GW100 project which studied a set of
molecules with different codes, each of which had its own distinct approaches to the $GW$ problem.\cite{GW100}

We have implemented the spinor $GW$/BSE approach in BerkeleyGW in order to provide an independent implementation of this method
for the general improvement of methodology in this area. This work also allows calculations in BerkeleyGW which
is a widely used and well established code in the community, with extensive testing. BerkeleyGW also has particular
advantages for $GW$/BSE with respect to massively parallel performance \cite{Mauro2}, Coulomb truncation and
interpolation \cite{BerkeleyGW}, and sampling schemes for reduced-dimensional systems \cite{nonuniform}.
In this paper, we present the results of this long-running implementation effort \cite{barker_thesis}, with a detailed
exposition of the formalism and in particular the handling of the effect of symmetries on spinors,
which has not been explicitly addressed in previous literature on spinor $GW$/BSE.
We also make careful comparisons to other codes, with their somewhat different technical details, to establish
the level of agreement achieved among spinor $GW$ calculations, and demonstrate the performance of spinor $GW$
vs. ``scalar-relativistic'' (``SR'') $GW$ calculations, in which only relativistic mass and Darwin terms are included in the construction of pseudopotentials, with SOC then included perturabitvely. While we make several direct comparisons for quasiparticle energy gaps and spin-orbit splittings for various conventional test systems such as Group IV and III-V semiconductors with BerkeleyGW and other two-component spinor $GW$ codes, the best comparison that is available for two-component spinor $GW$/BSE calculations for GaAs and GaSb is with Empirical Pseudopotential Method calculations incorporating spin-orbit coupling perturbatively\cite{cohen_epm_soc}. 

The ability to use two-component spinors in $GW$ and $GW$/BSE scientific software allows for the study of magnetic phenomena in materials, beyond the usual single-axis spin-polarized calculations of self-energy corrections for majority and minority spin channels in materials such as bulk Fe and NiO$_2$\cite{bgw_spin_polarized,aguilera_fe,bgw_nio,ag_nio}. 
While spin susceptibilities have been approximated within the usual spin-polarized $GW$ method\cite{zhu_overhauser,kukkonen_overhauser,lischner_sf},
Ref. \cite{biermann} derives results from many-body perturbation theory for susceptibilities describing spin-spin and spin-charge interactions. Spin susceptibilities\cite{friedrich2010spin} may then be used to calculate electron-magnon contributions to quasiparticle energies, as in the recent work in Ref. \cite{electron-magnon}. Other codes do not seem to have this functionality implemented.
To assist in non-collinear or antiferromagnetic calculations, magnetic symmetry groups have been exploited in the code Yambo to reduce the necessary size of magnetic systems to the primitive chemical unit cell\cite{yambo19}. While BerkeleyGW can treat magnetic systems within the supercell approach, the inclusion of spin susceptibilities, as well as the use of magnetic symmetry groups, is an ongoing work. We consider test systems with no magnetization in the present work.

This paper is structured as follows.
In Section \ref{sec:theory}, we review the theory of one-particle and two-particle
excited states within many-body perturbation theory in the $GW$ approximation,
and how such calculations are performed in a plane-wave basis with wavefunctions
that have two spinor components, and we discuss the appropriate treatment of crystal symmetries with spinorial wavefunctions, via quaternions,
in a plane-wave basis set. In Section \ref{sec:test_systems}, we demonstrate
the accuracy of our implementation in the BerkeleyGW
software package with calculations of the quasiparticle bandstructures
and absorption spectra of materials containing small, moderate, and large
spin-orbit coupling strength, finding agreement within 10 meV for energy gaps compared to results from other codes -- with the exception of $\Bi2Se3$, a difficult case needing a more sophisticated treatment. In Section \ref{sec:comparison}, we compare to available results
from other fully-relativistic $GW$ codes
\cite{spex,west,yambo,turbomole,fhiaims,questaal_thesis,gpaw_soc,vasp}.
In Section \ref{sec:performance}, we discuss the additional expense of computations that use spinor wavefunctions.
In Section \ref{sec:conclusion}, we conclude and give an outlook for future development in spinor $GW$/BSE.


\section{Spinor Wavefunctions in Many-Body Perturbation Theory}
\label{sec:theory}

%

We begin by generalizing the $GW$/BSE formalism to spinors, following the non-spinor approach used in BerkeleyGW \cite{BerkeleyGW}. The derivation of the basic framework of Hedin's equations is presented in the Supplementary Materials \cite{suppmat}, in a simpler form than the more general derivation in Ref. \cite{biermann}.
The formalism starts with a mean-field solution (typically from Kohn-Sham DFT) from a non-collinear spin calculation\cite{theurich_hill}, in which spin is not a quantum number of the state (as in a spin-polarized or collinear calculation) but rather another argument of the wavefunction alongside $\rp$. The Kohn-Sham wavefunction $\phi^{\rm KS}_{n\kp}(\rp) = \sum_{\alpha=\uparrow , \downarrow} \phi^{\rm KS}_{n\kp\alpha}(\rp) |\alpha\rangle$, with $|\phi^{\rm KS}_{n\kp\uparrow}|^2 + |\phi^{\rm KS}_{n\kp\downarrow}|^2 = 1$, has the Kohn-Sham eigenvalue $\epsilon^{\rm KS}_{n\kp}$:
\begin{equation}
\sum_{\alpha,\beta} \int \d \rp \left(\phi^{\rm KS}_{n \kp \alpha}(\rp)\right)^{\dagger} H^{\rm KS}_{\alpha,\beta}\, \phi^{\rm KS}_{n \kp \beta}(\rp) = \epsilon^{\rm KS}_{n \kp}.
\end{equation}
To construct the one-particle Green's function in the frequency domain, we consider the quasiparticle wavefunctions $\phi_{n \kp}$ and energies $E_{n \kp}$, giving
\begin{equation}
G_{\alpha\beta}\left(\vr_1,\vr_2; \omega\right) = \sum_{n\kp} \frac{\phi_{n\kp\alpha}(\vr_1)\phi_{n\kp\beta}^*(\vr_2)}{\omega - E_{n\kp} - i\eta\,\textrm{sgn}(\mu - E_{n\kp})},
\end{equation}
%
where $\eta$ is some small positive constant and $\mu$ is the chemical potential.

In practice, we find that the Kohn-Sham wavefunctions $\phi^{\KS}_{n\kp}$
(with corresponding energy eigenvalues $\epsilon^{\KS}_{n\kp}$) are usually good approximations to the actual
quasiparticle wavefunctions\cite{hybertsen}, so we approximate the Green's function as
\begin{equation}
G_{\alpha\beta}\left(\vr_1,\vr_2; \omega\right) \approx \sum_{n\kp} \frac{\phi^{\KS}_{n\kp\alpha}(\vr_1){\phi^{\KS}}^{*}_{n\kp\beta}(\vr_2)}{\omega - E_{n\kp} -  i\eta\,\textrm{sgn}(\mu - E_{n\kp})}\, .
\end{equation}
%
We evaluate the quasiparticle energies with a one-shot ``$G_0W_0$'' procedure:
\begin{equation}
\label{eq:qpenergy}
E_{n\kp} = \epsilon_{n\kp} + \sum_{\alpha\beta} \langle n,\kp,\alpha | \Sigma_{\alpha\beta}\left(E_{n\kp}\right) - V^{\xc}\delta_{\alpha\beta} | n,\kp,\beta \rangle,
\end{equation}
with the Kohn-Sham orbitals being expressed in bra-ket notation. The electron-electron self-energy is given by $\Sigma_{\alpha\beta} = iG_{\alpha\beta}W$\cite{biermann,sakuma}, where $W$ is the screened Coulomb potential, and $V^{\xc}$ is the exchange-correlation potential. In this study, we consider only non-magnetic materials, for which $V^{\xc}$ has only spin-diagonal components. (A system with local magnetic moments has the spin-dependent exchange-correlation potential $V^{\xc}\delta_{\alpha\beta} + \vec{\sigma}_{\alpha\beta}\cdot \vec{B}^{\xc}$. See Ref.
\cite{bulik} and references therein for the full treatment of the exchange-correlation functional for magnetic systems, which needs to be considered when calculating quasiparticle energy corrections including electron-magnon scattering terms in the self-energy\cite{electron-magnon}.)

%

To determine $W$, we must first compute the polarizability $P$ \cite{hybertsen}:
\begin{align}
& W_{\Gp\Gp '}(\vq,\omega) = \epsilon_{\Gp \Gp '}^{-1}(\vq,\omega)v(\vq + \Gp'),\\
& \epsilon_{\Gp \Gp '}(\vq,\omega) = \delta_{\Gp\Gp '} - v(\vq + \Gp)P_{\Gp \Gp '}(\vq,\omega).
\end{align}

The polarizability matrix for real frequencies may be constructed from the Kohn-Sham eigenfunctions and eigenvalues as\cite{adler,wiser,catalin_thesis}
%
\begin{align*}
P_{\Gp\Gp'}(\vq,\omega) = \frac{1}{N_{\kp}}\sum\limits_{n}^{\textrm{occ}}\sum\limits_{n'}^{\textrm{emp}}\sum_{\kp} & M^{*}_{nn'}(\kp,\vq,\Gp)M_{nn'}(\kp,\vq,\Gp') \nn
& \left[\frac{1}{\epsilon_{n\kp+\vq} - \epsilon_{n'\kp}-\omega + i\eta}+ \frac{1}{\epsilon_{n\kp+\vq} - \epsilon_{n'\kp}+\omega + i\eta}\right],
\end{align*}
%
with $N_{\kp}$ being the number of $\kp$-points used to sample the Brillouin Zone, and $\vq$ the momentum transfer.
The matrix elements
\begin{equation}
M_{nn'}(\kp,\vq,\Gp) = \sum_{\alpha}\langle n,\kp+\vq,\, \alpha|e^{i(\vq+\Gp)\cdot \vr} |n',\kp,\alpha \rangle
\end{equation}
may be computed for all $\Gp$ by multiplying the Fourier transforms of the wavefunctions, for a spin component $\alpha$ common to both wavefunctions;
computing the inverse Fourier transform of this product\cite{BerkeleyGW,fleszar}; and then summing over spin index:
\begin{equation}
\label{eq:matrix_el}
M_{nn'}(\kp,\vq,\{\Gp\}) = \sum_{\alpha}\textrm{FFT}^{-1}\left(\phi^{*}_{n\kp+\vq\alpha}(\vr)\phi_{n'\kp\alpha}(\vr)\right).
\end{equation}
Since the (non-magnetic\cite{bulik}) polarizability has its physical origin from density fluctuations arising from the spin-independent Coulomb interaction,
the form of the polarizability
is identical to the case in which spin-orbit is neglected,
apart from the sums over the spin index in the computation of the matrix elements in Eq. \ref{eq:matrix_el},
the doubled number of both valence and conduction bands within the summation over basis states,
and any differences in eigenfunctions and eigenvalues.
In many cases, these differences are sufficiently small such that one may calculate the polarizability
using the Kohn-Sham eigenfunctions and eigenvalues from a scalar-relativistic DFT calculation\cite{sakuma,aguilera}. However, in this work, we use the Kohn-Sham eigenfunctions and eigenvalues from fully-relativistic DFT calculations (``FR-DFT'').


%
%



We note that self-energy operator $\Sigma = iGW$ inherits the spin-dependence from the Green's function\cite{biermann},
but the process of taking matrix elements
reduces this spin-dependence to computing traces over the spinor components of the wavefunctions.
This is readily seen by considering the matrix elements of $\Sigma$, which separate into two terms,
with the screened-exchange (``SX'') coming from the poles of $G$, and the Coulomb-hole (``COH'') coming from the poles of $W$\cite{hybertsen}.
The matrix elements
$\langle n,\kp | \Sigma^{\SX} (\omega)|n,\kp \rangle$ and $\langle n,\kp | \Sigma^{\COH} (\omega)|n,\kp \rangle$
are given by
\begin{align}
\label{eq:sigma_sx}
\sum_{\alpha,\beta}\langle n,\kp,\alpha | \Sigma^{\SX}_{\alpha\beta}(\omega)|n,\kp,\beta\rangle = -\sum\limits_{n''}^{\textrm{occ}} \sum\limits_{\vq\Gp\Gp'} & M^{*}_{n''n}(\kp,-\vq,-\Gp)M_{n''n}(\kp,-\vq,-\Gp') \nn
& \times \epsilon_{\Gp\Gp'}^{-1}(\vq;\omega-E_{n''\kp-\vq})v(\vq+\Gp'),
\end{align}
\begin{align}
\label{eq:sigma_coh}
\sum_{\alpha,\beta}\langle n,\kp,\alpha | \Sigma^{\COH}_{\alpha\beta}(\omega)|n,\kp,\beta\rangle = -\sum\limits_{n''} \sum\limits_{\vq\Gp\Gp'} & M^{*}_{n''n}(\kp,-\vq,-\Gp)M_{n''n}(\kp,-\vq,-\Gp') \nn
& \times \frac{1}{\pi }\int \d \omega ' \frac{ \textrm{Im}\, \epsilon_{\Gp,\Gp '}^{-1}(\vq,\omega ') }{\omega - \epsilon_{n''\kp} - \omega ' + i\eta}.
\end{align}

To simplify the calculation of matrix elements of $\Sigma^{\COH}$,
a generalized plasmon pole model (GPP)\cite{complex_gpp,hybertsen} may be used.
 In this case, only the static dielectric function needs to be explicitly computed.
The Hybertsen-Louie GPP\cite{hybertsen} is justified through the use of a sum rule that is derived
from a double-commutator of a one-particle non-interacting Hamiltonian (as in the RPA) with charge density operators, and is therefore independent of spin-orbit coupling.

To calculate the optical absorption spectrum of a material, we may first try to evaluate the
imaginary part of the macroscopic dielectric function within the independent-particle approximation.
We may readily determine, using the usual expression\cite{yu_cardona,cohen_louie},
the imaginary part of the dielectric function to be
\begin{equation}
\varepsilon_2 (\omega) = \frac{8\pi^2e^2}{\omega^2}\sum_{vc\kp}\left|\bm{\lambda}\cdot \sum_{\alpha \beta}\langle v,\kp,\alpha| \vv_{\alpha \beta} | c,\kp,\beta\rangle\right|^2 \nonumber\delta\left(\omega - \left(E_{c\kp}-E_{v\kp}\right)\right),
\end{equation}
where $\bm{\lambda}$ is the direction of light polarization. The velocity operator $\vv = i \left[ H, \rp \right]$ now has a spin-dependence inherited from the Hamiltonian. However, it can be transformed just as in the spin-independent case (explained in Ref. \cite{BGW_arXiV}, containing a few additional details or corrections compared to Ref. \cite{BerkeleyGW}) into
\begin{equation}
\langle v,\kp | \vv | c,\kp \rangle = -i \left( E_{c\kp} - E_{v\kp} \right) \langle v,\kp | \rp | c,\kp \rangle
\label{eq:dipole}
\end{equation}
containing now a spin-independent dipole operator, and with an energy that cancels the $\omega^{-2}$ factor. We evaluate in practice:
\begin{equation}
\varepsilon_2 (\omega) = 8\pi^2e^2 \sum_{vc\kp} \left|\bm{\lambda}\cdot \sum_{\alpha}\langle v,\kp,\alpha| \rp | c,\kp,\alpha\rangle\right|^2 \nonumber\delta\left(\omega - \left(E_{c\kp}-E_{v\kp}\right)\right)
\end{equation}
where the dipole matrix element is calculated via a $\vq \rightarrow 0$ limit.
The momentum operator $-i \nabla$ can be used to approximate $\vv$ to avoid needing a set of wavefunctions on a shifted $k$-grid, but this is a worse approximation than in the spinless case, as the fully relativistic Hamiltonian contains additional non-local terms, not only the spin-orbit coupling but also both of the scalar relativistic terms \cite{hamann_1989}.

More accurate calculations of the absorption spectrum require the inclusion of excitonic effects,
which require a treatment of the many-body physics of an optically excited electron interacting with a hole. Since this interaction is Coulombic and therefore does not carry an intrinsic dependence on spin, the derivation of the Hamiltonian for this interaction within many-body perturbation theory, the Bethe-Salpeter Equation, can be expected to yield the same Hamiltonian as for the spinless case, only with traces over the spin indices in the matrix elements defined in Ref. \cite{rohlfing}.

Ref. \cite{yambo_bse_soc} presents a derivation of the Bethe-Salpeter Equation with two-component spinor wavefunctions, based on a treatment and notation for many-body perturbation theory from Ref. \onlinecite{stefanucci2013nonequilibrium}. We, however, wish to present a derivation that employs the Schwinger approach to many-body perturbation theory as in Ref. \cite{strinati}, in which a small perturbing electrostatic potential $\Phi$ is introduced to arrive at the relation between the (interacting) electron-hole propagator $L$ and the functional derivative of the one-particle Green's function with respect to this potential, following the work and notation conventions in Ref. \onlinecite{strinati}. As in Ref. \onlinecite{yambo_bse_soc}, we only consider inclusion of electron-electron interactions, unlike Ref. \cite{biermann} as we are not computing $GT$ or $GWT$ contributions to the self-energy\cite{electron-magnon}.

As an outline for our approach to the derivation of the BSE with two-component spinor wavefunctions, we begin with a definition of the interacting electron-hole propagator $L$, demonstrate its relation to a functional derivative of the one-particle Green's function within the Schwinger approach, and arrive at a Dyson series for $L$ from this functional derivative, which then allows us to identify the electron-hole interaction kernel, $K$. From the poles of the inverted Dyson series, we arrive at the BSE, in a basis set of electronic transitions from occupied bands to unoccupied bands, each composed of two-component spinors.

The electron-hole propagator $L$, is defined from the one-particle Green's function $G$ and the two-particle Green's function $G^{(2)}$, by
\begin{equation}
\label{eq:L2}
L_{\alpha\beta\gamma\zeta}(12',21') \equiv G_{\alpha\gamma} (12) G_{\beta\zeta} (2'1') - G^{(2)}_{\alpha\beta\gamma\zeta}(12',21').
\end{equation}
where 1 is an abbreviation for $\mathbf{r}_1, t_1$, and the same for 1', 2, and 2'. ``1$^{+}$'' is an abbreviation for $\mathbf{r}_1$, $t_1 + \eta$, with $\eta$ a small positive number.
We can introduce the non-interacting propagator as
\begin{equation}
L^0_{\alpha\beta\gamma\zeta}(12',21') = G_{\alpha\zeta}(11')G_{\beta\gamma}(2'2),
\end{equation}
which simply describes a pair of non-interacting particles.

To arrive at a Dyson-like equation for $L$ and to eliminate $G^{(2)}$, we introduce a spatially non-local electrostatic perturbing potential $\Phi$ to generate the appropriate number of coordinates for a four-point function such as $L$.
This introduces the additional perturbing term in the Hamiltonian, $ H^{'}(t) = \sum_{\beta\zeta}\int \d 2 \d 3\; \fod_{\beta}(2^+) \Phi(23)\delta(t-t_2)\delta(t - t_3)\delta_{\beta\zeta} \fo_{\zeta}(3)$, with $\fod_{\beta}(2^+)$ and $\fo_{\zeta}(3)$ fermionic field operators.

In the interaction picture, the perturbing term yields the time-development operator
\begin{equation}
\Sop = \exp\left\{-i \left(\int^{+\infty}_{-\infty} \d t \; H^{'}(t) \right)\right\},
\end{equation}
allowing us to rewrite the (single-particle) Green's function as
\begin{equation}
G_{\alpha\gamma}(11') = -i\; \frac{\langle \Psi^N_0 | \hat{T} \left[ \Sop \fo_{\alpha}(1) \fod_{\gamma}(1')\right] | \Psi^N_0 \rangle}{\langle \Psi^N_0 | \hat{T} \left[ \Sop \right] | \Psi^N_0 \rangle}\, ,
\end{equation}
where $|\Psi^N_0 \rangle$ is the $N$-particle ground-state wavefunction and $\hat{T}$ is the time-ordering operator.
A variation in the perturbing potential, $\delta \Phi$, will create a variation in the Green's function
\begin{align}
\label{eq:varg}
\delta G_{\alpha\gamma}(11') & = (-i)^2\sum_{\beta}\int\d 2 3\;  \delta\Phi(23)\;  \frac{\langle \Psi^N_0 | \hat{T} \left[ \Sop \fod_{\beta}(2^+) \fo_{\beta}(3) \fo_{\alpha}(1) \fod_{\gamma}(1')\right] | \Psi^N_0\rangle}{\langle \Psi^N_0 | \hat{T} \left[ \Sop \right] | \Psi^N_0 \rangle} \nn
&-(-i)G_{\alpha\gamma}(11') \sum_{\beta}\int\d 2 3\; \delta\Psi(23)\;  \frac{\langle \Psi^N_0 | \hat{T} \left[ \Sop \fod_{\beta}(2^+) \fo_{\beta}(3) \right] | \Psi^N_0\rangle}{\langle \Psi^N_0 | \hat{T} \left[ \Sop \right] | \Psi^N_0 \rangle}.
\end{align}

Recognizing that the first term on the right-hand side in Eq. \ref{eq:varg} is a two-particle Green's function $G^{(2)}$, apart from the multiplication by the $\delta\Phi$ and the integration over its coordinates, we have the variational derivative
\begin{equation}
\label{eq:nonlocvar}
\frac{\delta G_{\alpha\gamma}(11')}{\delta \Phi(23)} = \sum_{\beta}\left(-G^{(2)}_{\alpha\beta\gamma\beta}(13,1'2^+) + G_{\alpha\gamma}(11') G_{\beta\beta}(32^+)\right).
\end{equation}
 We note that in the context of non-collinear spins, this result was first derived in Ref. \cite{biermann}, though now generalized to include non-local perturbations. (We note that time-reversal symmetry-breaking magnetic perturbations can be included in the previous derivation; this is done in Ref. \cite{karlsson}.)



To define the electron-hole interaction kernel $K$, we find the Dyson series for $L$, using Eqs. \ref{eq:L2} and \ref{eq:nonlocvar}, and make use of the inverted Dyson series for $G$ ($G^{-1} = G^{(0){-1}} - \Sigma$):
\begin{align}
\label{eq:bse_dyson}
\sum_{\beta} L_{\alpha\beta\gamma\beta}(12',21') & = \frac{\delta G_{\alpha\gamma}(12)}{\delta \Phi(1'2')} \nn
& = -\sum_{\zeta\iota}\int\d 34\; G_{\alpha\zeta}(13)\frac{\delta G^{-1}_{\zeta\iota}(34)}{\delta \Phi(1'2')}G_{\iota\gamma}(42) \nn
& = - \sum_{\zeta\iota}\int \d 34\; G_{\alpha\zeta}(13)\frac{\delta (G^{(0)-1}_{\zeta\iota}(34) - \Sigma_{\zeta\iota}(34))}{\delta \Phi(1'2')}G_{\iota\gamma}(42) \nn
& = \sum_{\beta} G_{\alpha\beta}(11')G_{\beta\gamma}(2'2) + \sum_{\zeta\iota}\int \d 34\; G_{\alpha\zeta}(13)\frac{\delta \Sigma_{\zeta\iota}(34)}{\delta \Phi(1'2')}G_{\iota\gamma}(42) \nn
& = \sum_{\beta}L^{(0)}_{\alpha\beta\gamma\beta}(12',21') \nn
&\;\; + \sum_{\zeta\iota\mu\nu}\int\d 3456\; G_{\alpha\zeta}(13)\frac{\delta \Sigma_{\zeta\iota}(34)}{\delta G_{\mu\nu}(65)}\frac{\delta G_{\mu\nu}(65)}{\delta \Phi(1'2')}G_{\iota\gamma}(42)\nn
& = \sum_{\beta}L^{(0)}_{\alpha\beta\gamma\beta}(12',21')\nn
&\;\; + \sum_{\zeta\iota\mu\nu\beta}\int\d 3456\; G_{\alpha\zeta}(13)\frac{\delta \Sigma_{\zeta\iota}(34)}{\delta G_{\mu\nu}(65)}L_{\mu\beta\nu\beta}(62',51')G_{\iota\gamma}(42)\nn
& = \sum_{\beta}L^{(0)}_{\alpha\beta\gamma\beta}(12',21')\nn
&\;\; + \sum_{\zeta\iota\mu\nu\beta}\int\d 3456\; L^{(0)}_{\alpha\iota\gamma\zeta}(14,23)\frac{\delta \Sigma_{\zeta\iota}(34)}{\delta G_{\mu\nu}(65)}L_{\mu\beta\nu\beta}(62',51')\, .\nn
\end{align}

Having arrived at the Dyson series for $L$, we define the electron-hole interaction kernel $K$ to write the inverse Dyson equation as $L^{-1}_{\alpha\beta\gamma\beta}(12',21') = L^{(0)-1}_{\alpha\beta\gamma\beta}(12',21') -K_{\alpha\beta\gamma\beta}(12',21')$.

The electron-hole interaction kernel $K$ is determined from
\begin{align}
K_{\zeta\nu\iota\mu}(35,46) & = \frac{\delta \Sigma_{\zeta\iota}(34)}{\delta G_{\mu\nu}(65)} =\frac{\delta \left( v(37)\rho(7)\delta(34)\delta_{\zeta\iota} + iG_{\zeta\iota}(34)W(3^{+}4)\right) }{\delta G_{\mu\nu}(65)}\nn
& = -i \delta(34)\delta_{\zeta\iota}v(37)\frac{\delta G_{\rho\rho}(77^+)}{\delta G_{\mu\nu}(65)}\nn
&\;\; \; + i W(3^+4)\frac{\delta G_{\zeta\iota}(34)}{\delta G_{\mu\nu}(65)} \nn
& = -i \delta(34)\delta_{\zeta\iota}v(37)\delta(67)\delta(57^+)\delta_{\mu\rho}\delta_{\nu\rho} \nn
& \;\;\; + i W(3^+4) \delta(36)\delta(45)\delta_{\zeta\mu}\delta_{\iota\nu} \nn
& = -i v(36)\delta(34)\delta(56^+)\delta_{\zeta\iota}\delta_{\mu\nu} \nn
& \;\;\; + i W(3^+4) \delta(36)\delta(45)\delta_{\zeta\mu}\delta_{\iota\nu}.
\end{align}

In the above, we make the approximation that
$\frac{\delta \left( G_{\zeta\iota}(34)W(3^{+}4) \right)}{\delta G_{\mu\nu}(65)} \approx W(3^{+}4)\frac{\delta G_{\zeta\iota}(34)}{\delta G_{\mu\nu}(65)}$,
which simplifies the kernel and is found to be adequate in practice\cite{rohlfing}.





We wish to solve for the poles of $L^{-1}$, which requires us to decide on a basis for our excited states. In the basis of Kohn-Sham wavefunctions, the two-particle excited-state wavefunction \cite{onida_reining_rubio} for state $S$ is
\begin{equation}
\Psi^S_{\alpha\beta}(11') = \sum_{vc\vk} A^S_{vc\vk}\phi_{c\vk\alpha}(1)\phi^{*}_{v\vk\beta}(1')
+ B^S_{vc\vk} \phi_{v\vk\alpha}(1) \phi^{*}_{c\vk\beta}(1').
\end{equation}
In the Tamm-Dancoff Approximation (TDA), we take $B^S = 0$ and simplify the wavefunction:
\begin{equation}
\Psi^{S,(\textrm{TDA})}_{\alpha\beta}(11') = \sum_{vc\vk} A^S_{vc\vk}\phi_{c\vk\alpha}(1)\phi^{*}_{v\vk\beta}(1').
\end{equation}
The following uses exclusively the TDA to simplify the construction of the results; however the process may be generalized to include the off-diagonal blocks of the BSE allowed by non-zero $B^S$.

In the orbital basis, then, we obtain the same results for the inverted Dyson equation as in Ref.  \cite{rohlfing},
%
\begin{align}
& L^{(0)-1}_{\alpha\beta\gamma\beta}(12';21';\omega) =  \sum_{vc\vk} M^{\alpha\beta}_{cv\vk}(11')\left(\omega - E_{cv\vk}\right){M^{\beta\gamma}_{cv\vk}}^{*}(2'2) -M^{\gamma\beta}_{vc\vk}(22')\left(\omega + E_{cv\vk}\right){M^{\beta\alpha}_{vc\vk}}^{*}(1'1)\nn
& L^{-1}_{\alpha\beta\gamma\beta}(12',21';\omega) = \sum_{S}\left[ \Psi^S_{\alpha\beta}(11')\left(\omega - \Omega^S \right){\Psi^S_{\beta\gamma}}^{*}(2'2)) -\Psi^S_{\gamma\beta}(22')\left(\omega+\Omega^S\right){\Psi^S_{\beta\alpha}}^{*}(1'1) \right],
\end{align}
with $M^{\alpha\beta}_{vc\vk}(11')=\phi_{c\vk\alpha}(1)\phi^{*}_{v\vk\beta}(1')$ and $ E_{cv\vk} = E_{c\vk} - E_{v\vk}.$

We solve the inverted Dyson equation, enforcing the solution $\omega =\Omega_S$ and using orthonormality of the excited-state wavefunctions, to arrive at the Bethe-Salpeter Equation within the Tamm-Dancoff Approximation:
%

\begin{equation}
\left(E_{c\vk} - E_{v\vk}\right) A^S_{vc\vk} 
+ \sum_{v'c'\vk '}\left( K^{\textrm{X}}_{vc\kp,v'c'\kp '} + K^{\textrm{D}}_{vc\kp,v'c'\kp'}\right) A^S_{v'c'\vk '} \nn
 = \Omega^S A^S_{vc\vk} \, ,
\end{equation}
for the excited state $S$, with energy $\Omega_S$ and envelope function $A^S_{vc\vk}$.

This is exactly the form of the eigenvalue equation from the BSE when neglecting spin-orbit coupling, with the exception that we have twice as many valence and conduction states. 


The matrix elements of $K^X$ and $K^D$ are
\begin{align}
& K^{\textrm{X}}_{vc\kp,v'c'\kp '}  = -i \sum_{\alpha , \beta}\int\d 12\; \psi^*_{c'\vk'\alpha}(1)\psi_{v'\vk'\alpha}(1) v(12) \psi_{c\vk\beta}(2)\psi^*_{v\vk\beta}(2),\nn
& K^{\textrm{D}}_{vc\kp,v'c'\kp '} = i \sum_{\alpha , \beta}\int\d 12\; \psi^*_{c'\vk'\alpha}(1)\psi_{c\vk\alpha}(1) W(34) \psi_{v'\vk'\beta}(2)\psi^*_{v\vk\beta}(2).
\end{align}
The spin-dependence of the kernel, and thus the BSE, reduces as expected to a trace over the spin coordinates in the computation of the relevant generalized charge density matrix element:
\begin{equation}
M_{nn'}(\kp,\vq,\Gp) = \sum_{\alpha} \langle n,\kp+\vq,\, \alpha |e^{i(\vq+\Gp)\cdot \vr} |n',\kp, \alpha \rangle.
\end{equation}

The absorption spectrum, including the interacting two-component spinor electron and hole wavefunctions, is then computed from
\begin{equation}
\epsilon_2 (\omega) = 8\pi^2e^2 \sum_{S}\,\left|\,\sum_{vc\kp} A^S_{vc\kp}
\, \bm{\lambda}\cdot \sum_{\alpha}\langle \,\kp,\alpha| \rp | c,\kp,\alpha\rangle\right|^2 \nonumber\delta\left(\omega - \Omega^S \right),
\end{equation}
using the excitonic version of Eq. \ref{eq:dipole} \cite{BGW_arXiV}.

We also note that in the presence of spin-orbit coupling, spin is generally no longer a good quantum number,
so it is no longer possible to refactor the Bethe-Salpeter Hamiltonian into spin-singlet and -triplet block-diagonal submatrices\cite{rohlfing}.
Further, the number of valence and conduction bands both double, relative to spinless calculations. This makes explicit diagonalization
of the BSE Hamiltonian, which scales as $N^3_{\textrm{basis}} = (N_v N_c)^3$, more expensive by a factor of 64 for solids, as the basis has quadrupled. However, the time spent performing this diagonalization and computation of the absorption spectra remains a relatively rapid calculation, compared to calculation of the screened interaction and the self-energy (see Sec. \ref{sec:performance}).

The above derivation holds even when the Tamm-Dancoff Approximation is not invoked, since, again,
the screened and bare Coulomb interactions do not depend on spin.
Metals and semi-metals, generally speaking, and sometimes other systems\cite{beyond_tda}, have absorption spectra calculated with the full BSE Hamiltonian, with the Tamm-Dancoff Approximation\cite{nonTDA}.
As the BerkeleyGW routines for calculating unrestricted BSE use the matrix element calculation routine that is compatible
with spinor wavefunctions, this functionality in BerkeleyGW is also compatible with spinor wavefunctions, though
no such calculations are performed in this work.

The most formidable computational challenge with the inclusion of two-component spinor wavefunctions is the increase in the number of charge-density matrix elements (Eq. \ref{eq:matrix_el}, which must be calculated for the polarizability, self-energy, and BSE kernel). Compared to a calculation performed on the same system without spin, the number of both valence and conduction states double. Taking the ratio of the scaling of the charge-density matrix element calculation with system size $N$ \cite{BerkeleyGW}, we find an increase in computation time by
\begin{equation}
\frac{(2 N)^2\, 2 \log (2N)}{N^2\log N} = 8(1+\log_{N}2),
\end{equation}
where the additional factor of 2 in the numerator comes from having to compute the inverse-FFT for each of the of two-component spinor wavefunctions. Since, at best, we are increasing the cost of matrix element calculations by more than a factor of 8, we should make use of symmetries of the Brillouin Zone to allow for converged calculations within reasonable computational cost. (Detailed discussion about the performance of the major sections of the BerkeleyGW code is included in Section \ref{sec:performance}.)

\section{Spinor wavefunctions and symmetries}

\subsection{Symmetries, No spin}

We first briefly review the use of symmetries for wavefunctions with plane-wave basis functions in the absence of spin\cite{BerkeleyGW}, as this is also necessary when using spinor wavefunctions. We use the symmetries of the Brillouin Zone to store only the necessary wavefunction coefficients within
the irreducible wedge.
When computing
the polarizability and self-energy, we then unfold to
``little group of the $q$-vector''
except when computing off-diagonal matrix elements of the self-energy\cite{BerkeleyGW}.

We consider the action of a transformation $\rot$, a proper or improper rotation, followed by a fractional translation $\btau$ on a Bloch wavefunction
$\psi_{n\kp}(\rp) = u_{n\kp}(\rp)e^{i\kp\cdot \rp} = e^{i\kp\cdot \rp} \sum_{\Gp} c_{n\kp}(\Gp) e^{i\Gp\cdot\rp}$\cite{tinkham1964group}.
A coordinate vector
$\rp$ is transformed as $\rp' = \rot \rp +\btau$, and the Bloch wavefunction is transformed according to
\begin{align}
P_{\{\rot|\btau\}} \psi_{n\kp}(\rp) & = \psi_{n\kp}(\rot^{-1}\rp - \rot^{-1}\btau) \nn
& = u_{n\kp}(\rot^{-1}\rp - \rot^{-1}\btau) e^{i\kp\cdot\left(\rot^{-1}\rp-\rot^{-1}\btau\right)} \nn
& = u_{n\kp}(\rot^{-1}\rp - \rot^{-1}\btau)e^{i \rot\kp\cdot \left(\rp-\btau\right)}.
\end{align}
where we use the property that scalars formed in dot products are invariant under rotation of both vectors. We drop the phase factor $e^{-i\rot\kp\cdot\btau}$, as this is common among all bands at a given k-point.

We expand the periodic part of the Bloch function in its plane-wave basis:
\begin{align}
u_{n\kp}\left(\rot^{-1}\rp - \rot^{-1}\btau\right) & = \sum_{\Gp}  c_{n\kp}(\Gp) e^{i\Gp\cdot\left(\rot^{-1}\rp - \rot^{-1}\btau \right)} \nn
& = \sum_{\rot^{-1}\Gp} c_{n\kp}(\rot^{-1}\Gp) e^{-i\Gp\cdot \btau}e^{i\Gp\cdot \rp}.
\end{align}
Substituting in to the previous equation and reordering the summation over G-vectors,
\begin{align}
& P_{\{\rot|\btau\}} \psi_{n\kp}(\rp) = \tilde{u}_{n\rot\kp}(\rp) e^{i \rot\kp\cdot \rp},\nn
& \tilde{u}_{n\rot\kp}(\rp) = \sum_{\Gp} \tilde{c}_{n\rot\kp}(\rot^{-1}\Gp) e^{-i\Gp\cdot \btau} e^{i\Gp\cdot \rp}.
\end{align}
(The use of the $\tilde{c}$ refers to the change in the function $u$ when evaluated at $\rp$ instead of $\rot^{-1}\rp - \rot^{-1}\btau$.)

This allows us to use the usual relation when using symmetries to unfold the Brillouin Zone from an irreducible wedge:
\begin{equation}
c_{n\kp}(\Gp) \rightarrow \tilde{c}_{n\rot\kp}(\rot^{-1}\Gp)e^{-i\Gp\cdot\btau}.
\end{equation}

\subsubsection{Symmetries, with spinor wavefunctions}

We now extend the above discussion for the case of spinor wavefunctions, where
\begin{equation}
\psi_{n\kp}(\rp) = u_{n\kp\uparrow}(\rp) e^{i\kp\cdot \rp}\chi_{\uparrow}+u_{n\kp\downarrow}(\rp) e^{i\kp\cdot \rp}\chi_{\downarrow}.
\end{equation}
and $\chi_\alpha$ represents a spinor.
The periodic functions $u_{n\kp\uparrow}$ and $u_{n\kp\downarrow}$ are spatial and thus transform according to the above. However, the spinor itself
rotates according to the rules of transformation for elements of the group $\textrm{SU}(2)$:
\begin{align}
P_{\{\rot|\btau\}}\psi_{n\kp}(\rp) & = \tilde{u}_{n\rot\kp\uparrow}(\rp) \exp{(i\rot\kp\cdot \rp)} \exp{(i  \,\hat{n}\cdot \vec{\sigma}\, \theta / 2)} \chi_{\uparrow} \nn
& \; +\tilde{u}_{n\rot\kp\downarrow}(\rp) \exp{(i\rot\kp\cdot\rp)}\exp{(-i\, \hat{n}\cdot \vec{\sigma}\, \theta / 2)}\chi_{\downarrow}\, ,
\end{align}
where $\hat{n}$ and $\theta$ are the unit-axis and angle (about the axis $\hat{n}$) that recreates the rotational action of the symmetry operation $\rot$.
We readily arrive at the rule for transforming two-component spinor Bloch functions:
\begin{equation}
P_{\{\rot|\btau\}}\psi_{n\kp}(\rp) = 
\left(
\begin{matrix}
\cos\left(\frac{\theta}{2}\right) -in_z\sin\left(\frac{\theta}{2}\right) &  
(-n_y-in_x)\sin\left(\frac{\theta}{2}\right) \\
(-n_y+in_x)\sin\left(\frac{\theta}{2}\right) &
\cos\left(\frac{\theta}{2}\right)+in_z\sin\left(\frac{\theta}{2}\right) 
\end{matrix}
\right)
\left(
\begin{matrix}
\tilde{u}_{n\rot\kp\uparrow}(\rp) e^{i\rot\kp\cdot \rp} \\
\tilde{u}_{n\rot\kp\downarrow}(\rp) e^{i\rot\kp\cdot \rp}
\end{matrix}
\right),
\end{equation}
where $n_{\ii}$
is the $\textrm{i}$'th Cartesian component of $\hat{n}$.

The task, then, is to determine $\hat{n}$ and $\theta$ for each symmetry $\rot$ used in unfolding the Brillouin Zone.

\subsubsection{Axis and Angle Extraction}

The Euler axis/angle form of the matrix in the Cartesian basis that rotates a vector (in $\mathbb{R}^3$)
by some angle $\theta$ about a unit vector $\hat{n}$ can be recreated by consideration of the generators of the Lie group for $\textrm{SO}(3)$. We simply restate the end result\cite{markley_1978}:
\begin{equation}
\rot\left(\hat{n},\theta \right) = 
\left(\begin{matrix}
\cos\theta + \nx^2 \left(1-\cos\theta\right) & -\nz \sin\theta + \nx\ny \left(1 - \cos\theta\right) & \ny \sin\theta + \nx\nz \left(1-\cos\theta\right)\\
\nz \sin\theta + \ny\nx \left(1 - \cos\theta \right) & \cos\theta + \ny^2 \left(1-\cos\theta \right) & -\nx \sin\theta + \ny\nz\left(1-\cos\theta\right)\\
-\ny \sin\theta + \nz\nx\left(1-\cos\theta\right) & \nx \sin\theta + \nz\ny\left(1-\cos\theta\right) & \cos\theta + \nz^2 \left(1-\cos\theta \right)
\end{matrix}\right) \, .
\end{equation}

A first attempt to extract the values of $\nx$, $\ny$, $\nz$, and $\theta$ for any given rotation matrix might make use of the following:
\begin{align}
\cos\theta & = \frac{1}{2}\left( Tr\left(\rot\right) - 1 \right),\nn
n_{\textrm{k}} & = \epsilon_{\textrm{ijk}} \left(-\left[ \rot \right]_{\textrm{ij}} + \left[ \rot \right]_{\textrm{ji}} \right)/2\sin\theta,
\end{align}
with $\epsilon_{\textrm{ijk}}$ being the Levi-Civita tensor. However, the division by $\sin\theta$ is singular for $\theta = 0 \pm \eta$ or $\theta = \pi \pm \eta$, where $\eta$ is
the machine precision for floating-point numbers. Further, if $Tr\left(\rot\right) = -1 - 2\eta$, the use of the inverse-cosine function to find $\theta$
is also subject to failure.

More general extraction algorithms attempt to remove singularities by making use of a reparameterization of the above rotation matrix in terms of a four-component quaternion.
A unit quaternion $q$, with transpose $q^T = (q_1,\, q_2 ,\, q_3, \, q_4)$,  can be parameterized by $\hat{n}$ and $\theta$ from
\begin{equation}
q^T =
(n_1 \sin \frac{\theta}{2},\,
n_2 \sin \frac{\theta}{2},\,
n_3 \sin \frac{\theta}{2},\,
\cos \frac{\theta}{2}).
\end{equation}
We now rewrite the rotation matrix above as\cite{markley_1978}
\begin{equation}
\rot =
\left(\begin{matrix}
q^2_1 - q^2_2 - q^2_3 + q^2_4  & 2\left(q_1q_2 + q_3q_4 \right)  & 2\left(q_1q_3 - q_2q_4\right)  \\
2\left(q_2q_1 - q_3q_4 \right)  & -q^2_1 + q^2 - q^2_3 + q^2_4  & 2\left(q_2q_3+q_1q_4\right)  \\
2\left(q_3q_1 + q_2q_4\right)  & 2\left(q_3q_2 - q_1q_4\right)  & -q^2_1 -q^2_2 + q^2_3 + q^2_4
\end{matrix}\right).
\end{equation}

We extract the parameters $\hat{n}$ and $\theta$ by using Markley's\cite{markley} modification to Shepperd's algorithm\cite{shepperd}.
We first construct an auxiliary matrix $\mathbf{X}$:
\begin{equation}
\mathbf{X} = 
\left(
\begin{matrix}
1 + 2\rot_{11} - Tr\left(\rot\right) & \rot_{21} + \rot_{12}               & \rot_{31} + \rot_{12}               & \rot_{23} - \rot_{32} \\
\rot_{12} + \rot_{21}               & 1 + 2\rot_{22} - Tr\left(\rot\right) & \rot_{32} + \rot_{23}               & \rot_{31} - \rot_{13} \\
\rot_{13} + \rot_{31}               & \rot_{23} + \rot_{32}               & 1 + 2\rot_{33} - Tr\left(\rot\right) & \rot_{12} - \rot_{21} \\
\rot_{23} - \rot_{32}               & \rot_{31} - \rot_{13}               & \rot_{12} - \rot_{21}                 & 1 + Tr\left(\rot\right)
\end{matrix}\right).
\end{equation}
Then we compute the norms of each column $\mathbf{x}^{\textrm{i}}$. We use the column with the largest norm to compute the quaternion $q$,
from
\begin{equation}
q = \pm \mathbf{x}^{\textrm{i}}/\left|\mathbf{x}^{\textrm{i}} \right|,
\end{equation}
by construction\cite{markley}.
The positive and negative branches of solution come from the double-cover of $\textrm{SO}(3)$ by $\textrm{SU}(2)$, the latter of which
is parameterized by the four real quaternion components $q^{\textrm{i}}$ instead of the usual two complex components for spin.
While recently developed modifications to Shepperd's algorithm allow for continuous quaternions through the use of solutions from both the positive and negative branches\cite{wu2019optimal}, we arbitrarily choose to use the positive branch, which is adequate for materials science applications, as this algorithm produces errors that are bounded to the order of round-off error\cite{markley}.

With a quaternion $q$ that is now guaranteed to be non-singular, we evaluate the rotation angle $\theta$ from
\begin{equation}
\theta = 2\arctan\left(\frac{q^2_1+q^2_2+q^2_3}{q_4} \right),
\end{equation}
where $\arctan$ is a function with all real numbers as its domain,
and evaluate the $i$'th component of the axis of rotation $n_i$ from
\begin{equation}
n_i = \frac{q_i}{\sqrt{q^2_1+q^2_2+q^2_3}}.
\end{equation}


The set of rotation matrices $\rot$ for a crystalline system are usually stored in the basis of lattice vectors in $ab$ $initio$ codes,
as it allows these matrices (up to 48 in number) to be written with nine integers. In this case, we must transform the rotation matrices
in the lattice basis, $\rot^{\textrm{lat}}$, to the rotation matrix in the Cartesian basis. If we form a matrix $A$ out of the three
lattice vectors $\mathbf{a}_1$, $\mathbf{a}_2$, and $\mathbf{a}_3$ as
\begin{equation}
A = 
\left(\begin{matrix}
a_{1x} & a_{2x} & a_{3x} \\
a_{1y} & a_{2y} & a_{3y} \\
a_{1z} & a_{2z} & a_{3z}
\end{matrix}\right),
\end{equation}
this transformation is
\begin{equation}
\rot^{\textrm{cart}} = A \rot^{\textrm{lat}} A^{-1}.
\end{equation}
If instead we decide to use the reciprocal lattice vectors $\mathbf{b}_1$, $\mathbf{b}_2$, $\mathbf{b}_3$ to construct the matrix $B$ in a fashion as in the above,
we make use of the identity $B^{\textrm{T}}A = 2\pi I$ to write
\begin{equation}
\rot^{\textrm{cart}} = \left(B^{\textrm{T}}\right)^{-1} \rot^{\textrm{lat}} B^{\textrm{T}}.
\end{equation}
This latter choice is beneficial if the matrices $A$ and $B$ are in fact stored as their transposes, as some codes do.

Finally, we note that in the presence of inversion symmetries, ``improper rotations'' $S$ must be considered. While often considered to be the composition of a rotation and a mirror reflection about the plane perpendicular to the axis of the rotation, instead we can consider the improper rotation $S$ to be (in general, a different) rotation $\rot$ followed by inversion $N$,  $S = N\rot$\cite{tinkham1964group}. 
However, if both spatial inversion and time-reversal operations commute with the Hamiltonian under consideration,
the (spinor) wavefunction is a simultaneous eigenstate of both symmetries. Thus in the presence
of only time-reversal symmetric terms in the Hamiltonian, the wavefunctions
are unaffected by inversion, apart from perhaps an overall phase factor.
We identify improper rotations by the identity $\textrm{det}\left( S \right) = -1$, and if detected, use only the rotation part $R$ of $S$ to transform the spinor components of the wavefunction.

\section{Test systems}
\label{sec:test_systems}

We present results for seven different materials with a wide range of spin-orbit coupling (SOC) strengths. The diamond and zincblende semiconductors
Si, Ge, and GaAs are technologically important materials with weak SOC. GaSb has a spin-orbit splitting of its valence bands of similar magnitude
as its band gap. CdSe has a wurtzite structure and a significant SOC (429~{meV}\cite{hellwege}, over 25 times larger than that of wurtzite GaN, 16.8~{meV}\cite{madelung}).
Au is a prototypical metal with strong SOC. Finally, Bi$_2$Se$_3$ has a nontrivial topological nature due to the band inversion induced by its strong SOC,
and is a particularly challenging case to explore which has been studied in much previous literature.

\subsection{Computational Details}

We compute mean-field wavefunctions and eigenvalues from Density Functional Theory\cite{kohn_sham,hohenberg_kohn}.
For the exchange-correlation energy, we employ the Perdew-Zunger parameterization of the LDA\cite{perdew_zunger}.
We generate fully-relativistic 
pseudopotentials using the Optimized Norm-Conserving Vanderbilt Pseudopotential (ONCVPSP) scheme\cite{oncvpsp}
with parameters from the Pseudo-Dojo pseudopotential database\cite{pseudo_dojo}.
The pseudopotentials for Au, Bi, Cd, Ga, Ge, and Sb contain the full shell of the semicore states (e.g., $5s^25p^65d^{10}$ for Bi)
for accurate calculation of the bare exchange\cite{semicore}.
All DFT calculations are carried out with the Quantum ESPRESSO software package\cite{espresso}.


We first determine the equilibrium lattice constants and atom positions.
Table \ref{tab:parameters} shows that all relaxed lattice constants are in very good agreement with experimental measurements.
We instead use the experimental lattice parameters and atomic coordinates for $\Bi2Se3$ due to the sensitivity of its DFT bandstructure with respect to its geometry{\cite{functional_sensitivity}.


Next, the quasiparticle energies are computed with the one-shot ``$G_0W_0$'' approach,
using the Hybertsen-Louie Generalized Plasmon Pole model\cite{hybertsen,complex_gpp} for the inverse dielectric matrix.
For the case of bulk Au, we also
calculated the quasiparticle band structure in the Godby-Needs Plasmon Pole Model\cite{godby_needs} and found differences of 50~{meV}
or smaller in the quasiparticle energies, in the range 6~{eV} above and below the Fermi energy.
Table \ref{tab:parameters} summarizes our parameters for the empty state summations, the k-point sampling, and the plane-wave cutoffs for the dielectric matrices.
We use the static remainder method to improve convergence with the number of empty states
in the Coulomb-hole summation\cite{static_remainder}.
We verified that $G_0W_0$ evaluation
of the self energy in the band-diagonal approximation yields quantitatively accurate bandstructures for these test systems. Deviations from this methodology in the computation of the bandstructure for $\Bi2Se3$ are enumerated in Section \ref{sec:bi2se3}.

The k-point sampling and number of bands used in constructing the $GW$-BSE Hamiltonian are summarized in Table \ref{tab:bse_params}.
All excited-state calculations are carried out with the BerkeleyGW software package.

%
%

\begin{table}[p]
\caption{\label{tab:parameters} The kinetic energy cutoffs $E_{\textrm{cut}}$, calculated lattice parameters, experimental lattice parameters,  Brillouin zone sampling, screened Coulomb cutoff $\epsilon_{\textrm{cut}}$, and number of empty states used in the sums
for both the polarizability (``Chi'') and the Coulomb-hole (``COH'') term in the self-energy. The pseudopotentials for Ge, Sb, Cd, and Au contain the full shell of the semicore states (e.g., $4s^24p^64d^{10}$ for Sb)\cite{semicore}. The experimental lattice parameters are from Ref. \cite{madelung}. 
For Si, Ge, GaP, GaAs, and GaSb, we use the same parameters as Ref. \cite{malone}, and for Au, Ref. \cite{jamal}. GaP results are discussed in Section \ref{sec:comparison} .}
\begin{ruledtabular}
  \begin{tabular}{lcccccc}
  & $E_{\textrm{cut}}$ (Ry) & $a^{\textrm{relaxed}}_0(\mbox{\AA})$ & $a^{\textrm{exp.}}_0(\mbox{\AA})$ & $\kp$-grid & $\epsilon_{\textrm{cut}}$ (Ry) & Empty States \\
  \hline
  Si   & 120 & 5.48 & 5.47 & 8$\times$8$\times$8 & 20 & 800 \\
  Ge   & 120 & 5.63 & 5.66 & 8$\times$8$\times$8 & 25 & 600 Chi, 1000 COH \\
  GaP  & 350 & 5.45 & 5.45 & 8$\times$8$\times$8 & 40 & 800 Chi, 1000 COH \\
  GaAs & 350 & 5.61 & 5.65 & 8$\times$8$\times$8 & 20 & 1002 \\
  GaSb & 350 & 6.09 & 6.10 & 8$\times$8$\times$8 & 20 & 1002 \\
  CdSe & 200 & 4.30 & 4.30 & 6$\times$6$\times$4 & 20 & 996 \\
  Au   & 72 & 4.08 & 4.08 & 8$\times$8$\times$8 & 50 & 2018 \\
  \end{tabular}
  \end{ruledtabular}
\end{table}

\begin{table}[p]
\caption{\label{tab:bse_params}
The values of the Brillouin Zone sampling of the fine grid, the number of valence and conduction bands used as the basis for the BSE, and the Gaussian broadening of the delta function.}
\begin{ruledtabular}
  \begin{tabular}{lcccc}
  & $\kp_{\textrm{fine}}$ grid & $N_v$ & $N_c$ & Broadening (meV)\\
      \hline 
        GaAs & 12$\times$12$\times$12 & 6 & 8 & 150 \\
        GaSb & 12$\times$12$\times$12 & 6 & 8 & 100  \\
        Au   & 12$\times$12$\times$12 & 6 & 4 & 150 \\

  \end{tabular}
  \end{ruledtabular}
\end{table}

\subsection{Si}\label{sec:si}

\begin{figure}[p]
\centering
  \includegraphics[width=0.5\textwidth]{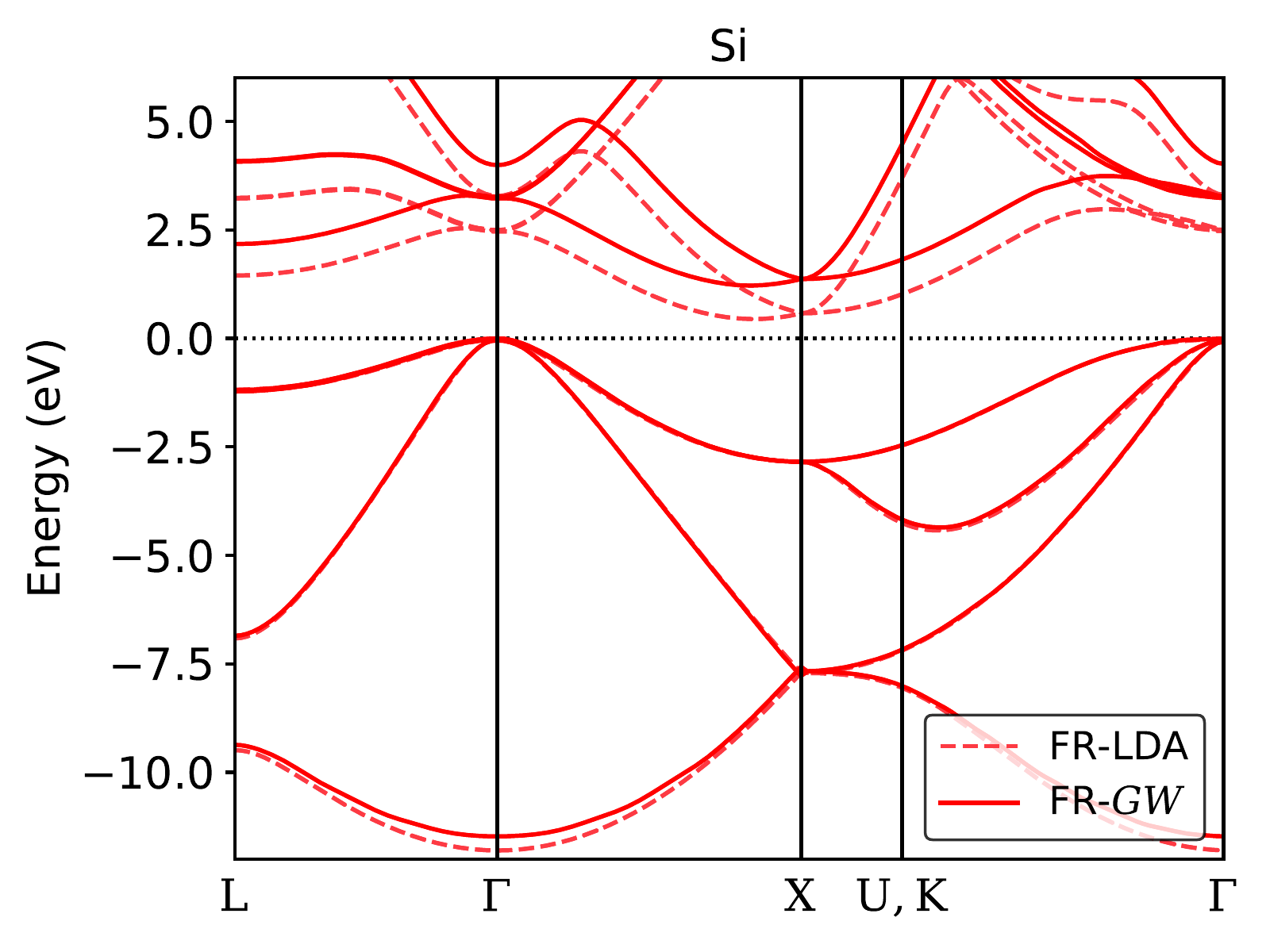}
  \caption{The electronic bandstructure of Si.  Fully-relativistic (``FR'') LDA and $GW$ in dashed and solid lines, respectively.
}
  \label{fig:Si}
\end{figure}

Fig. \ref{fig:Si} shows the bandstructure of Si calculated with fully-relativistic LDA (``FR-LDA'') and fully-relativistic $GW$ (``FR-GW'').
We find the FR-LDA band gap to be 0.445 eV, with the valence band maximum at the $\Gamma$-point and the conduction
band minimum along the $\Gamma - X$ line. The FR-$GW$ gap
is 1.22 eV, with the valence band maximum and conduction band minimum occuring at the same k-points as in FR-LDA.
The measured band gap is 1.17~{eV} at low-temperature; correcting for the zero-point electron-phonon renormalization (ZPR) yields
a gap of 1.22\cite{cardona_thewalt_zpr} or 1.23 eV\cite{cardona_zpr}, in excellent agreement with our FR-$GW$ result.
Table \ref{tab:si_soc} shows that the calculated spin-orbit splittings from FR-LDA and FR-$GW$ are in
excellent agreement with experiment, and calculation at the $GW$ level has little effect on these splittings. A comparison with a $GW$+SOC calculation\cite{malone} shows that the perturbative treatment of spin-orbit coupling gives good agreement with FR-$GW$ results for interband gaps and spin-orbit splittings, within few 10~{meV}.

In the absence of spin-orbit coupling (scalar relativistic, ``SR''), the band gaps increase slightly to 0.46~{eV} in LDA and 1.23~{eV} in $GW$.
The SR-$GW$ direct gap at $\Gamma$ is also slightly larger at 3.26~{eV}, compared to 3.22~{eV} within FR-$GW$.



\begin{table}[H]
\caption{
\label{tab:si_soc}
The band gap and spin-orbit splitting for Si, computed at the FR-LDA and FR-$GW$ levels, compared to experiment. The fundamental band gap from experiment is reported with Zero-Point Renormalization corrections.}
\begin{ruledtabular}
\begin{tabular}{lcccc}
 &  FR-LDA & FR-$GW$ & $GW$+SOC\cite{malone} &  Experiment \\
\hline
$E_g$ (eV) & 0.45 & 1.22 & 1.27 & 1.22 \cite{cardona_thewalt_zpr}, 1.23 \cite{cardona_zpr}\\
$E(\Gamma_{6c})-E(\Gamma_{8v})$ (eV) & 2.46 & 3.22 & 3.28 &  3.34 \cite{si_direct_gap}\\
$\Delta^{\SOC}(\Gamma,v)$ (eV) & 0.05 & 0.05 & 0.05 & 0.044 \cite{madelung}\\
$\Delta^{\SOC}(\Gamma,c)$ (eV) & 0.03 & 0.04 & 0.04 & 0.030 - 0.040 \cite{madelung} \\
$\Delta^{\SOC}(L,v)$ (eV)      & 0.03 & 0.03 & 0.03 & 0.030 \cite{madelung} \\
$\Delta^{\SOC}(L,c)$ (eV)      & 0.01 & 0.01 & 0.02 & -- \\
\end{tabular}
\end{ruledtabular}
\end{table}

\subsection{Ge}\label{sec:ge}

\begin{figure}[p]
\centering
  \includegraphics[width=0.5\textwidth]{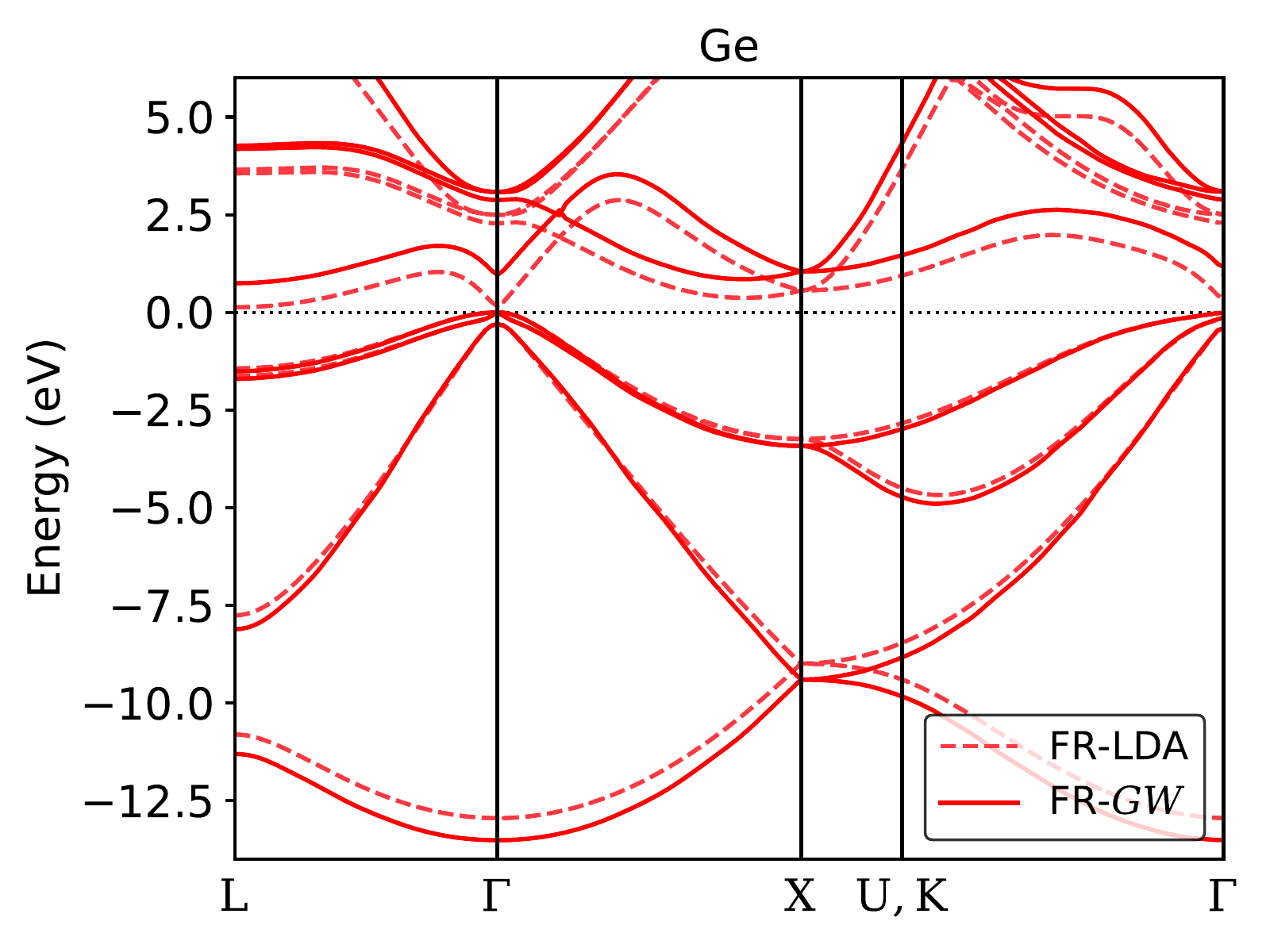}
  \caption{The electronic bandstructure of Ge. Fully-relativistic (``FR'') LDA and $GW$ in dashed and solid lines, respectively.
}
  \label{fig:Ge}
\end{figure}

Fig. \ref{fig:Ge} shows the bandstructure of Ge from FR-LDA and FR-$GW$.
DFT calculations often find a negative band gap for Ge\cite{bachelet,semicore,kotani,ku_eguiluz}. With the use of 
the fully-relativistic ONCVPSP pseudopotential with 3$s^2$3$p^6$3$d^{10}$ semicore states, however, we find a small but positive
direct gap at the $\Gamma$-point of 0.15~{eV} (Table \ref{tab:ge_soc}).

The FR-LDA indirect band gap is found to be 0.13~{eV}, with the valence band maximum at $\Gamma$ and the conduction band minimum at $L$.
Self-energy corrections at the FR-$GW$ level increase this gap to 0.743~{eV}.
The experimental gap is 0.744~{eV}\cite{madelung}, increasing to 0.79~{eV}\cite{cardona_zpr} when ZPR is taken into account, in good agreement
with the FR-$GW$ result.
Table \ref{tab:ge_soc} shows the calculated spin-orbit splittings. The splittings in FR-$GW$ are in better agreement with experimental data compared to that in FR-LDA by a few~{meV}. The perturbative treatment of SOC\cite{malone} has good agreement with the FR-$GW$ results for spin-orbit splittings. We attribute the underestimated band gaps from Ref. \cite{malone} from the use of a Ge pseudopotential that freezes the $n=3$ semicore states in the core, rather than to inherent limitations of the perturbative approach.


The indirect band gap in the SR-$GW$ approach is 0.842~{eV}, 0.1~{eV} larger than in FR-$GW$. The direct gap at $\Gamma$
in FR-$GW$ is calculated to be 0.960~{eV}, smaller than the SR-$GW$ direct gap result of 1.05~{eV}.
The discrepancies in both the indirect and direct gaps compared to FR-$GW$ are due to the moderately strong SOC in Ge.
While the change of these gaps upon the inclusion of SOC can be calculated by the inclusion of SOC as a perturbation to the quasiparticle Hamiltonian as in Ref. \cite{malone}, another approach to a perturbative treatment of SOC for standard Group IV or III-V semiconductors approximates the valence-band maximum as purely atomic (cationic, for compound semiconductors) $p$ states, which split due to spin-orbit coupling as in a free atom, with $p_{3/2}$ states shifting upward in energy by $\frac{1}{3} \Delta^{\SOC}$ and $p_{1/2}$ downward by $\frac{2}{3} \Delta^{\SOC}$ \cite{cardona_soc}. The FR-LDA value for the spin-orbit splitting of these bands can be used to approximate $\Delta^{\SOC}$, so then a $GW$ gap can be quickly approximated from using the SR-$GW$ gap $E_{g}^{\textrm{SR}-GW}$ and the FR-LDA spin-orbit splitting $\Delta_{\textrm{LDA}}^{\SOC}(\Gamma,v)$:
\begin{equation}
E_{g}^{\textrm{FR}-GW} \approx E_{g}^{\textrm{SR}-GW} - \frac{1}{3} \Delta_{\textrm{LDA}}^{\SOC}(\Gamma,v).
\end{equation}
This ``atomic SOC perturbation'' estimates a $GW$+SOC direct bandgap at $\Gamma$ of 0.95~{eV}, which agrees with the FR-$GW$ value (0.960~{eV}) within 10~{meV}.

\begin{table}[H]
\caption{
\label{tab:ge_soc}
The band gap and spin-orbit splittings for Ge, computed at the FR-LDA and FR-$GW$ levels, compared to experiment. Experimental data is from Ref. \cite{madelung} unless otherwise specified. }
\begin{ruledtabular}
\begin{tabular}{lcccc}
 &  FR-LDA & FR-$GW$ & $GW$+SOC\cite{malone} & Experiment \\
\hline
$E_g$ (eV)                    & 0.13 & 0.74 & 0.54 & 0.79 \\
$E(\Gamma_{7c}) - E(\Gamma_{8v})$ (eV) & 0.15 & 0.96 & 0.38 & 0.90 \\
$\Delta^{\SOC}(\Gamma,v)$ (eV) & 0.31 & 0.30 & 0.32 & 0.297 \\
$\Delta^{\SOC}(\Gamma,c)$ (eV) & 0.22 & 0.21 & 0.24 & 0.200 \\
$\Delta^{\SOC}(L,v)$ (eV)      & 0.19 & 0.19 & 0.20 & 0.228 \\
$\Delta^{\SOC}(L,c)$ (eV)      & 0.10 & 0.08 & 0.12 & --\\
\end{tabular}
\end{ruledtabular}
\end{table}

\subsection{GaAs}\label{sec:gaas}

\begin{figure}[H]
  \centering
  \includegraphics[width=0.5\textwidth]{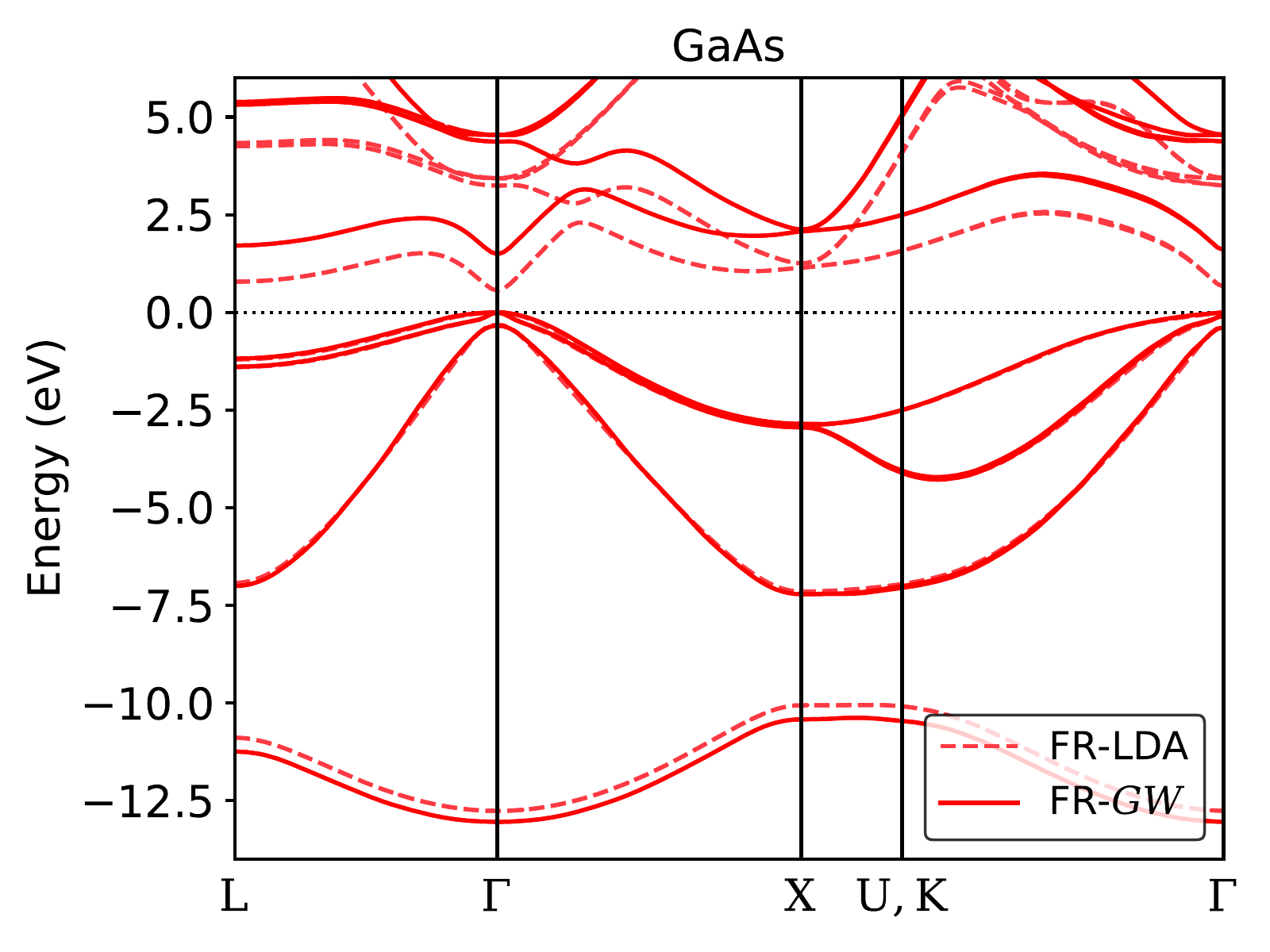}
  \caption{The electronic bandstructure of GaAs. Fully-relativistic (``FR'') LDA and $GW$ in dashed and solid lines, respectively.
}
  \label{fig:GaAs}
\end{figure}

\begin{figure}[H]
  \includegraphics[width=0.5\textwidth]{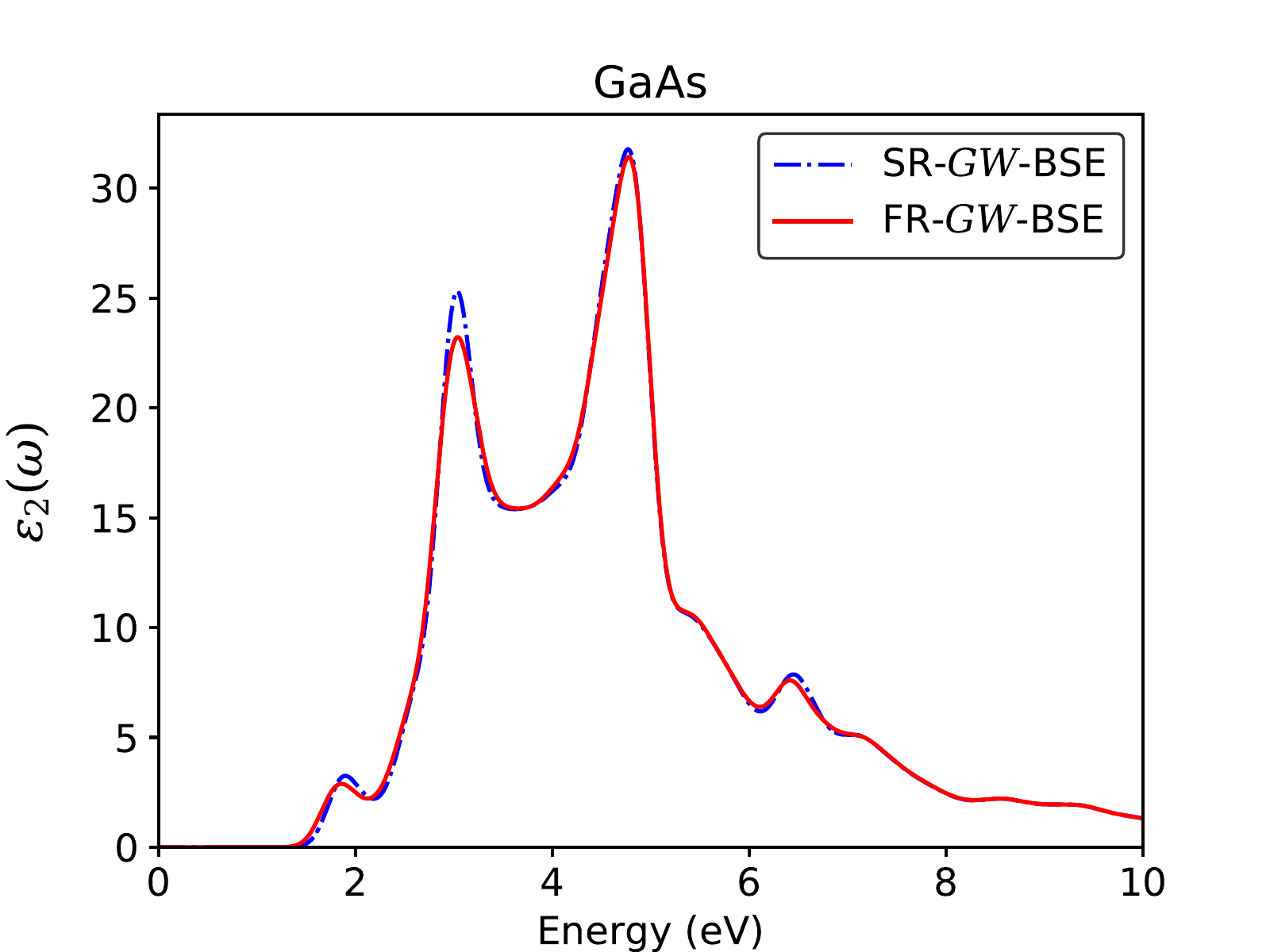}
  \caption{The absorption spectra of GaAs, calculated at the SR-$GW$-BSE (blue) and FR-$GW$-BSE (red) level.}
  \label{fig:figure_gaas_absp}
\end{figure}

Fig. \ref{fig:GaAs} shows the bandstructure for GaAs calculated with FR-LDA and FR-$GW$.
We find an FR-LDA band gap of 0.55~{eV} and a FR-$GW$ gap of 1.49~{eV}, compared to the 0 K gap of 1.52~{eV}, from experiment\cite{madelung}.
(Correcting for ZPR, the experimental band gap increases to 1.57~{eV}\cite{cardona_gaas}).
Table \ref{tab:gaas_so} shows the calculated spin-orbit splittings. We find that the FR-$GW$ splittings are in excellent agreement with experiment. The $GW$+SOC approach to the quasiparticle energies finds a band gap of 1.51~{eV}\cite{malone}, indicating that the perturbative approach is sufficient for GaAs. The ``atomic SOC perturbation'' approximation, based on a SR-$GW$ gap of 1.60~{eV}, estimates a $GW$+SOC gap of 1.49~{eV}.


Fig. \ref{fig:figure_gaas_absp} shows the absorption spectrum of GaAs from FR-$GW$-BSE and SR-$GW$-BSE.
Both methods yield similar spectra, apart from a small shift of 40~{meV} in the first peak after absorption onset, $E_0 + \Delta$, and a difference in amplitude of the $E_1$ absorption peak at 3~{eV}.
The shift in the $E_0 + \Delta$ peak is due to the smaller direct band gap within FR-$GW$ as compared to the SR-$GW$ result.
The $E_1$ peak has been measured to be split into a pair of peaks, $E_1$ and $E_1 + \Delta$ due to SOC, but the splitting (200~{meV})
is on the order of the resolution of the calculation (150~{meV}) with the given fine-grid k-point sampling of 12$\times$12$\times$12 and is thus obscured. The FR-$GW$-BSE absorption spectrum peak energies are compared with the Empirical Pseudopotential Method (``EPM'') results\cite{cohen_epm_soc} in Table \ref{tab:gaas_bse_epm} in which spin-orbit coupling is added as a perturbation to the interband transition energies (``EPM+SOC''), with the high level of agreement for the $E_0$ and $E_1$ peaks indicating the perturbative treatment of SOC is sufficient. The FR-$GW$-BSE for the $E_2$ peak energy, 4.77~{eV}, agrees less well with experiment (5.133~{eV} \cite{gaas_experiment_peaks}) than EPM+SOC (5.11~{eV} \cite{cohen_epm_soc}).

%

\begin{table}[H]
\begin{ruledtabular}
\caption{
\label{tab:gaas_so}
The band gap and spin-orbit splitting for GaAs, computed at the FR-LDA and FR-$GW$ levels, compared to experiment.
Experimental data is from Ref. \cite{madelung} unless otherwise specified.
}
\begin{tabular}{lcccc}

 &  FR-LDA & FR-$GW$ & $GW$+SOC\cite{malone} & Experiment \\
\hline
$E_g$ (eV)                     & 0.55  & 1.49  & 1.31 & 1.57\cite{cardona_gaas} \\
$\Delta^{\SOC}(\Gamma,v)$ (eV) & 0.32 & 0.34 & 0.35 & 0.340 \\
$\Delta^{\SOC}(\Gamma,c)$ (eV) & 0.19 & 0.17 & 0.20 & 0.171 \\
$\Delta^{\SOC}(L,v)$ (eV)      & 0.20 & 0.21 & 0.22 & 0.22 \\
$\Delta^{\SOC}(L,c)$ (eV)      & 0.08  & 0.07  & 0.09 & 0.05 \\

\end{tabular}
\end{ruledtabular}

\end{table}

\begin{table}
\caption{\label{tab:gaas_bse_epm}
Absorption peak energies for GaAs in eV. The energy for the spin-orbit split $E_1 + \Delta$ peak is not resolved in the present calculation or in Ref. \cite{gaas_experiment_peaks}.
}
\begin{ruledtabular}
\begin{tabular}{lccc}

 &  FR-$GW$-BSE & EPM+SOC\cite{cohen_epm_soc} & Experiment\cite{gaas_experiment_peaks} \\
\hline
$E_0 + \Delta$                & 1.85  & 1.86  & 1.851 \\
$E_1$                         & 3.03  & 3.03  &  3.041\\
$E_1 + \Delta$                & --  & 3.25  & -- \\
$E_2$                         & 4.77  & 5.11  & 5.133 \\
\end{tabular}
\end{ruledtabular}
\end{table}


\subsection{GaSb} \label{sec:gasb}

\begin{figure}[H]
  \centering
  \includegraphics[width=0.5\textwidth]{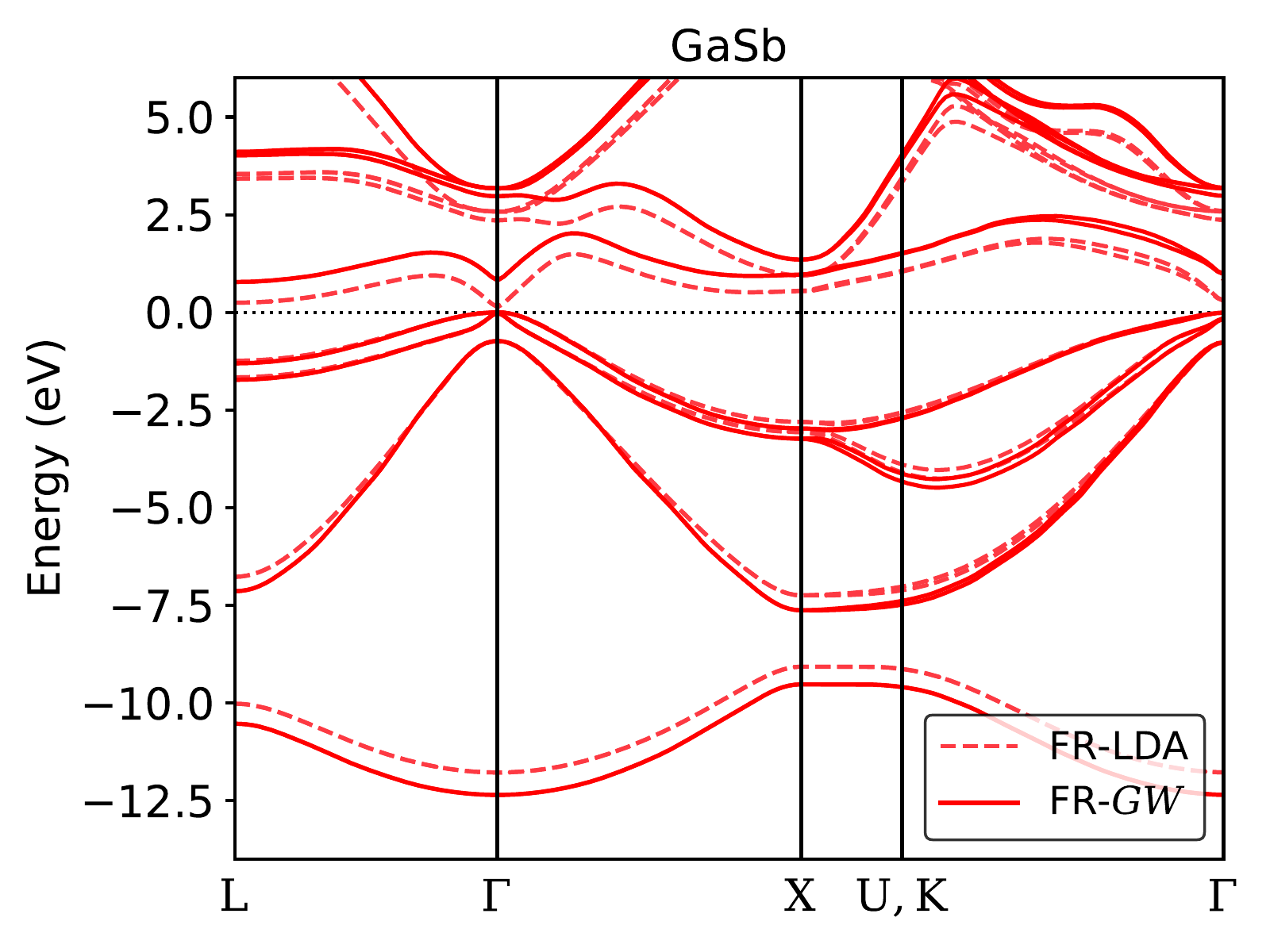}
  \caption{The electronic bandstructure of GaSb. Fully-relativistic (``FR'') LDA and $GW$ in dashed and solid lines, respectively.
}
  \label{fig:GaSb}
\end{figure}

\begin{figure}[H]
  \centering
  \includegraphics[width=0.5\textwidth]{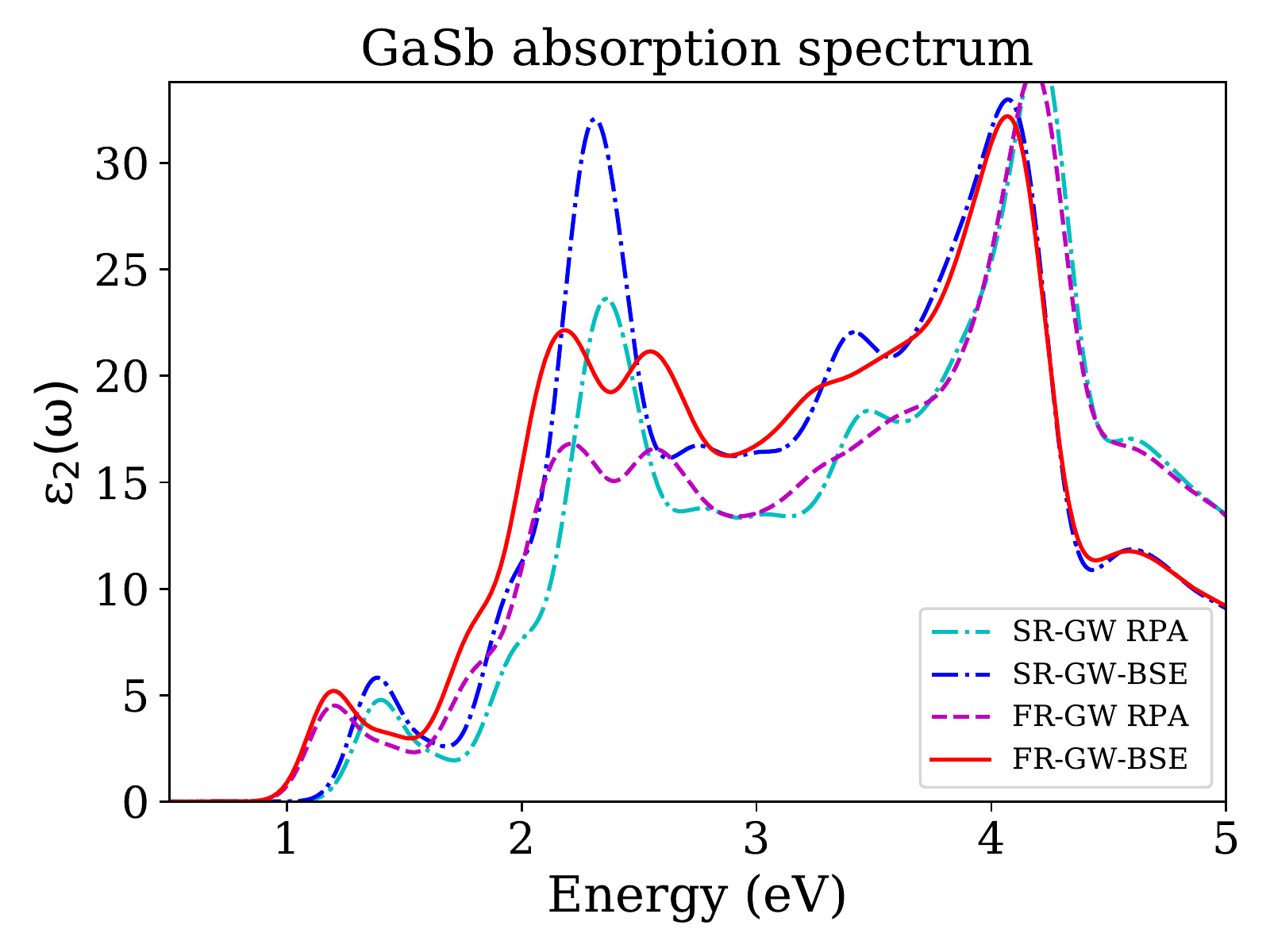}
  \caption{The absorption spectra of GaSb, calculated at the SR-$GW$-RPA (cyan), SR-$GW$-BSE (blue), FR-$GW$-RPA (magenta), and FR-$GW$-BSE (red) levels. RPA spectra are included to assess any renormalization of SOC by the electron-hole interaction.}
  \label{fig:figure_gasb_absp}
\end{figure}

Figure \ref{fig:GaSb} shows the FR-LDA and FR-$GW$ bandstructures.
We compute an FR-LDA band gap of 0.135~{eV} and a FR-$GW$ direct gap of 0.82~{eV},
compared to the low temperature gap of 0.82~{eV} from experiment\cite{madelung}.
Table \ref{tab:gasb_so} shows the calculated spin-orbit splittings, which are in good agreement with experiment.
However, at the FR-$GW$, as well as SR-$GW$, level, we find the conduction band minimum to be located at the $L$ point instead of the $\Gamma$-point,
despite experimental evidence of a direct band gap at the $\Gamma$-point in GaSb\cite{madelung}.

In the SR-LDA, the direct band gap at $\Gamma$ 
is 0.36~{eV}, and in the SR-$GW$, 1.07~{eV}.
As in FR-$GW$, the fundamental band gap is predicted to be indirect in SR-$GW$, from $\Gamma$ to $L$, with a value of 1.02~{eV}.
The direct gap is again well-approximated by applying the atomic perturbation theory estimate, 0.82~{eV}, while a rigorous $GW$+SOC perturbative treatment finds a direct band gap of 0.70~{eV}\cite{malone}.


Figure \ref{fig:figure_gasb_absp} shows the absorption spectrum for GaSb calculated with the SR-$GW$-BSE and FR-$GW$-BSE methods, as well as the non-interacting ``RPA'' method, in which the electron-hole kernel in the Bethe-Salpeter Equation is disregarded. The RPA spectra are included to assess any differences in the spin-orbit split peaks $E_1$ and $E_1 + \Delta$ due to renormalization of SOC from the electron-hole interaction.

The absorption spectrum of GaSb has significant differences when including SOC. First, the absorption onset is shifted
by 190 meV due to the large difference in the quasiparticle band gap when including (0.82~{eV}) or neglecting SOC (1.07~{eV}).
Also, we can clearly resolve the 2.3~{eV} peak splitting into the $E_1$ and $E_1 + \Delta$ peaks with the inclusion of SOC.
The $E_1$ and $E_1+\Delta$ peak placements at 2.18~{eV} and 2.54~{eV} agree well with the experimental\cite{cardona_gasb}
spectrum peak placements of 2.18~{eV} and 2.62~{eV}, respectively, and the EPM+SOC peak placements of 2.22~{eV} and 2.86~{eV}\cite{cohen_epm_soc}. These results, as well as the energies of the $E_0$ and $E_2$ peaks, are included in Table \ref{tab:GaSb_spectra_data}. The absorption spectra computed within RPA are qualitatively similar to that of the BSE, with the $E_1$ and $E_1 + \Delta$ peak splitting agreeing with that of the BSE under 10~{meV}, indicating no significant renormalization effects of SOC from the electron-hole interaction.

\begin{table}
\caption{\label{tab:GaSb_spectra_data}
Absorption peak energies for GaSb, in eV. The $E_0 + \Delta$ does not appear in Ref. \cite{cohen_epm_soc} or Ref. \cite{gasb_experiment_peaks}.
}
\begin{ruledtabular}
\begin{tabular}{lccc}

 &  FR-$GW$-BSE & EPM+SOC\cite{cohen_epm_soc} & Experiment\cite{gasb_experiment_peaks} \\
\hline
$E_0 + \Delta$                & 1.19  &  -- & -- \\
$E_1$                         & 2.18  & 2.22  & 2.184 \\
$E_1 + \Delta$                & 2.54  & 2.86  & 2.618 \\
$E_2$                         & 4.06  & 4.37  & 4.286 \\
\end{tabular}
\end{ruledtabular}
\end{table}

\begin{table}[H]
\begin{ruledtabular}
\caption{
\label{tab:gasb_so}
The band gap and spin-orbit splittings for GaSb, computed at the FR-LDA and FR-$GW$ levels, compared to experiment. Experimental data is from Ref. \cite{madelung}.}
\begin{tabular}{lcccc}
 &  FR-LDA & FR-$GW$ & $GW$+SOC\cite{malone} & Experiment \\
\hline
$E(\Gamma_{6c})-E(\Gamma_{8v})$ (eV) & 0.14  & 0.82  & 0.70 & 0.822 \\
$E(L_{6c})-E(\Gamma_{8v})$ (eV) & 0.25 & 0.78 & 0.85 & 0.907 \\
$\Delta^{\SOC}(\Gamma,v)$ (eV) & 0.74 & 0.73 & 0.73 & 0.756 \\
$\Delta^{\SOC}(\Gamma,c)$ (eV) & 0.23 & 0.20 & 0.21 & 0.213 \\
$\Delta^{\SOC}(L,v)$ (eV)      & 0.42 & 0.42 & 0.42 & 0.430 \\
$\Delta^{\SOC}(L,c)$ (eV)      & 0.12 & 0.09 & 0.12 & 0.13 \\

\end{tabular}
\end{ruledtabular}

\end{table}

\subsection{CdSe}

\begin{figure}[H]
  \centering
  \includegraphics[width=0.5\textwidth]{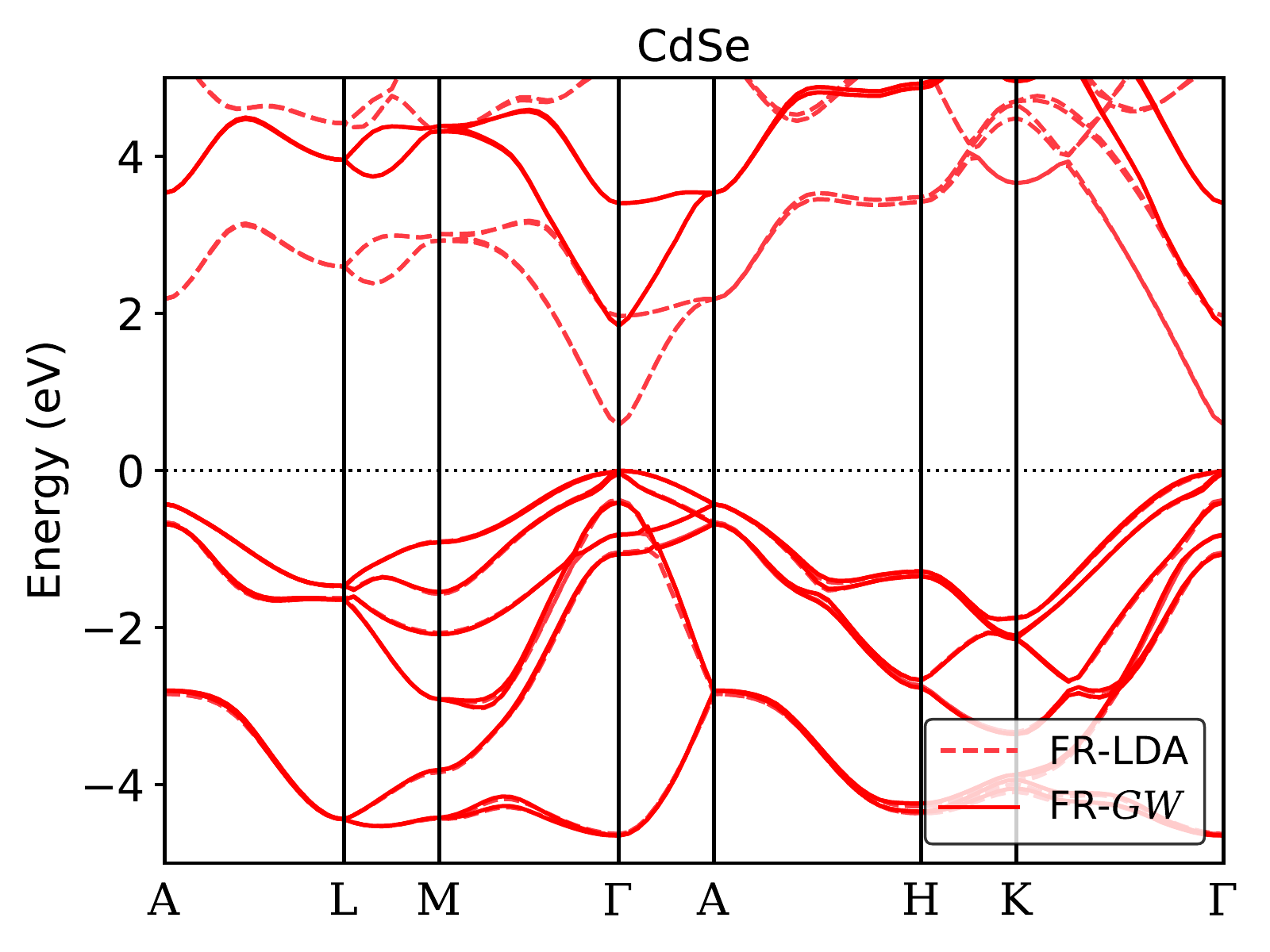}
  \caption{The electronic bandstructure of CdSe. Fully-relativistic (``FR'') LDA and $GW$ in dashed and solid lines, respectively.
}
  \label{fig:CdSe}
\end{figure}

Figure \ref{fig:CdSe} shows the bandstructure of CdSe computed at the FR-LDA and FR-$GW$ levels.
We compute the FR-LDA band gap of 0.58~{eV} and the FR-$GW$ gap of 1.85~{eV}, compared to the low temperature gap of 1.84~{eV} from experiment\cite{hellwege}. Since CdSe has a wurtzite lattice, its symmetry properties are different from the diamond and zincblende materials. Namely, the $\Gamma_1$ band is lower in energy than the top of the valence band, $\Gamma_6$, by the crystal-field splitting $\Delta^{\CF}$ \cite{yu_cardona}.
Table \ref{tab:cdse_eqp_data} shows the spin-orbit and crystal field splittings, with FR-$GW$ showing excellent agreement with experiment and much improved compared to FR-DFT.


The SR-$GW$ gap is larger due to the neglect of the large spin-orbit splitting, with a value of 1.99~{eV}. The change in the band gap due to the inclusion of spin-orbit coupling
is well-approximated by atomic perturbation theory, though the spin-orbit splitting differs by over 30~{meV} whether using LDA or $GW$ (Table \ref{tab:cdse_eqp_data}). The gap estimate is 1.86~{eV}, within a few tens of meV of the FR-$GW$ value, when using the
FR-LDA value of the spin-orbit splitting.

\begin{table}[H]
\begin{ruledtabular}
\caption{
\label{tab:cdse_eqp_data}
The band gap and spin-orbit splitting for CdSe, computed at the FR-LDA and FR-$GW$ levels, compared to experiment.
The spin-orbit (SOC) and crystal field (CF) splitting refers to the states at the top of the valence band at $\Gamma$.
Experimental data is from Ref. \cite{hellwege}.}
\begin{tabular}{lccc}

 &  FR-LDA & FR-$GW$ & Experiment \\
\hline
$E_g$ (eV)                    & 0.58 & 1.85 & 1.84 \\
$\Delta^{\SOC}(\Gamma,v)$ (eV) & 0.372 & 0.405 & 0.429 \\
$\Delta^{\CF}(\Gamma,v)$ (eV) & 0.036 & 0.026 & 0.026 \\

\end{tabular}
\end{ruledtabular}

\end{table}

\subsection{Au}

\begin{figure}[H]
  \centering
  \includegraphics[width=\textwidth]{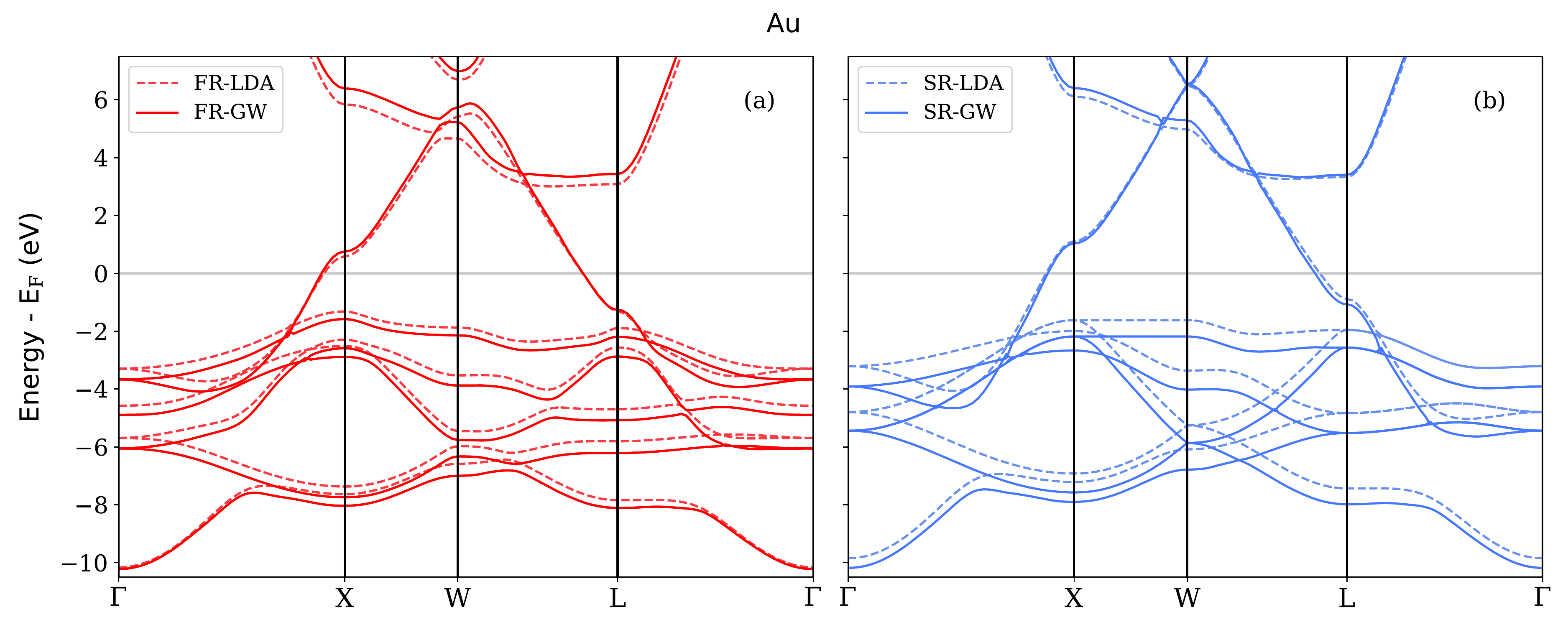}
  \caption{The electronic bandstructure of Au: (a) fully relativistic and (b) scalar relativistic, with LDA and $GW$ in dashed and solid lines, respectively.}
  \label{fig:Au}
\end{figure}

The bandstructure of Au computed from FR-LDA and FR-$GW$ is shown in Fig. \ref{fig:Au}a, and from SR-LDA and SR-$GW$ in Fig. \ref{fig:Au}b.
The inclusion of spin-orbit coupling changes the degeneracy of the occupied $s$-$d$ hybridized states, especially visible at the $\Gamma$, $W$, and $L$ high-symmetry points in the bandstructure.
The bandwidth for both valence and conduction bands increase upon inclusion of the electronic self-energy.
Table \ref{tab:Au_eqp_data} shows that the quasiparticle energies are generally improved with FR-$GW$ compared to FR-LDA,
especially near the Fermi level. The FR-$GW$ quasiparticle energies are in high agreement with a quasiparticle self-consistent $GW$ calculation in which SOC is added perturbatively, from Ref. \cite{tonatiuh}, indicating that the perturbative treatment of SOC for the bandstructure is sufficient. The Fermi level for FR-$GW$ and SR-$GW$ is recalculated using the the Bl\"ochl tetrahedron method\cite{tetrahedron} with the quasiparticle energies, from the \texttt{cms-py} Python library\cite{cmspy}.
The quasiparticle energies are largely similar whether using the Hybertsen-Louie or the Godby-Needs GPP model,
within an energy range of 6~{eV} above or below the Fermi level \cite{suppmat}.

The interband absorption spectrum for Au is shown in Fig. \ref{fig:figure_au_absp}, which shows a redshift in the onset of absorption with the inclusion of spin-orbit coupling, and an additional absorption peak at 1.6~{eV}, with the absorption spectra having minor qualitative changes when including the electron-hole interaction (``BSE'') or not (``RPA''). While the optical properties of Au are well-known to be impacted by relativistic effects\cite{romaniello_gold}, the inclusion of only scalar relativistic effects is insufficient for a description of its absorption of visible light. The absorption spectrum was calculated with a 12$\times$12$\times$12 k-point sampling, six valence bands, and four conduction bands.

\begin{table}
\begin{ruledtabular}
\caption{\label{tab:Au_eqp_data}The FR-LDA and FR-$GW$ band energies in eV for Au relative to the Fermi energy, as compared to QSGW+SOC and experiment. Bands at high-symmetry k-points are labelled according to their double-group irreducible representation (see Ref. \cite{tonatiuh}). }
\begin{tabular}{lccccc}

 &  FR-LDA & FR-$GW$ & QSGW+SOC\cite{tonatiuh} & Experiment \\
\hline

$\Gamma_{6}^{+}$ & -10.17 & -10.22 & -10.39 & --  \\
$\Gamma_{8}^{+}$ & -5.69  & -6.05  & -6.02 & -5.09\footnote{\label{Aua}Ref. \cite{gold_mills}}, -6\footnote{\label{Aub}Ref. \cite{gold_baalmann}}, -6.01\footnote{\label{Auc}Ref. \cite{gold_courths}}  \\
$\Gamma_{7}^{+}$ & -4.58  & -4.89 & -4.85 & -4.45\textsuperscript{\ref{Aua}}, -4.6\textsuperscript{\ref{Aub}}, -4.68\textsuperscript{\ref{Auc}}  \\
$\Gamma_{8}^{+}$ & -3.29  & -3.67  & -3.67 & -3.55\textsuperscript{\ref{Aua}}, -3.65\textsuperscript{\ref{Aub}}, -3.71\textsuperscript{\ref{Auc}}  \\
$\Gamma_{7}^{-}$ & 13.91  & 14.46 & 15.36 & 16\textsuperscript{\ref{Auc}}, 15.9\footnote{\label{Aud}Ref. \cite{gold_jaklevic_davis}}  \\
$\Gamma_{6}^{-}$ & 17.26  & 17.81 & 17.97 & 18.8\textsuperscript{\ref{Auc}}  \\
\hline
$L_{6}^{+}$      & -7.84  & -8.11  & -8.01 & -7.8\textsuperscript{\ref{Aub}}  \\
$L_{4,5}^{+}$    & -5.80  & -6.21  & -6.16 & -6.23\textsuperscript{\ref{Aub}}, -6.2\textsuperscript{\ref{Auc}}  \\
$L_{6}^{+}$      & -4.69  & -5.08  & -4.97 & -4.88\textsuperscript{\ref{Aub}}, -5\textsuperscript{\ref{Auc}}  \\
$L_{6}^{+}$      & -2.56  & -2.87  & -2.95 & -3.2\textsuperscript{\ref{Auc}}  \\
$L_{4,5}^{+}$    & -1.90  & -2.19  & -2.24 & -2.3\textsuperscript{\ref{Auc}}, -2.5\footnote{\label{Aue}Ref. \cite{gold_szczepanek}}  \\
$L_{6}^{-}$      & -1.32  & -1.26  & -1.63 & -1\textsuperscript{\ref{Aue}}, -1\footnote{\label{Auf}Ref. \cite{gold_jaklevic_lambe}}, -1.01\footnote{\label{Aug}Ref. \cite{gold_chen}}, -1.1\footnote{\label{Auh}Ref. \cite{gold_marel}}  \\
$L_{6}^{+}$      & 3.09   & 3.44  & 3.19 & 3.6\textsuperscript{\ref{Aue}}, 3.65\textsuperscript{\ref{Auf}}, 3.56\textsuperscript{\ref{Aug}}, 3.4\textsuperscript{\ref{Auh}} \\

\end{tabular}
\end{ruledtabular}

\end{table}


\begin{figure}[t]
  \centering
  \includegraphics[width=\linewidth]{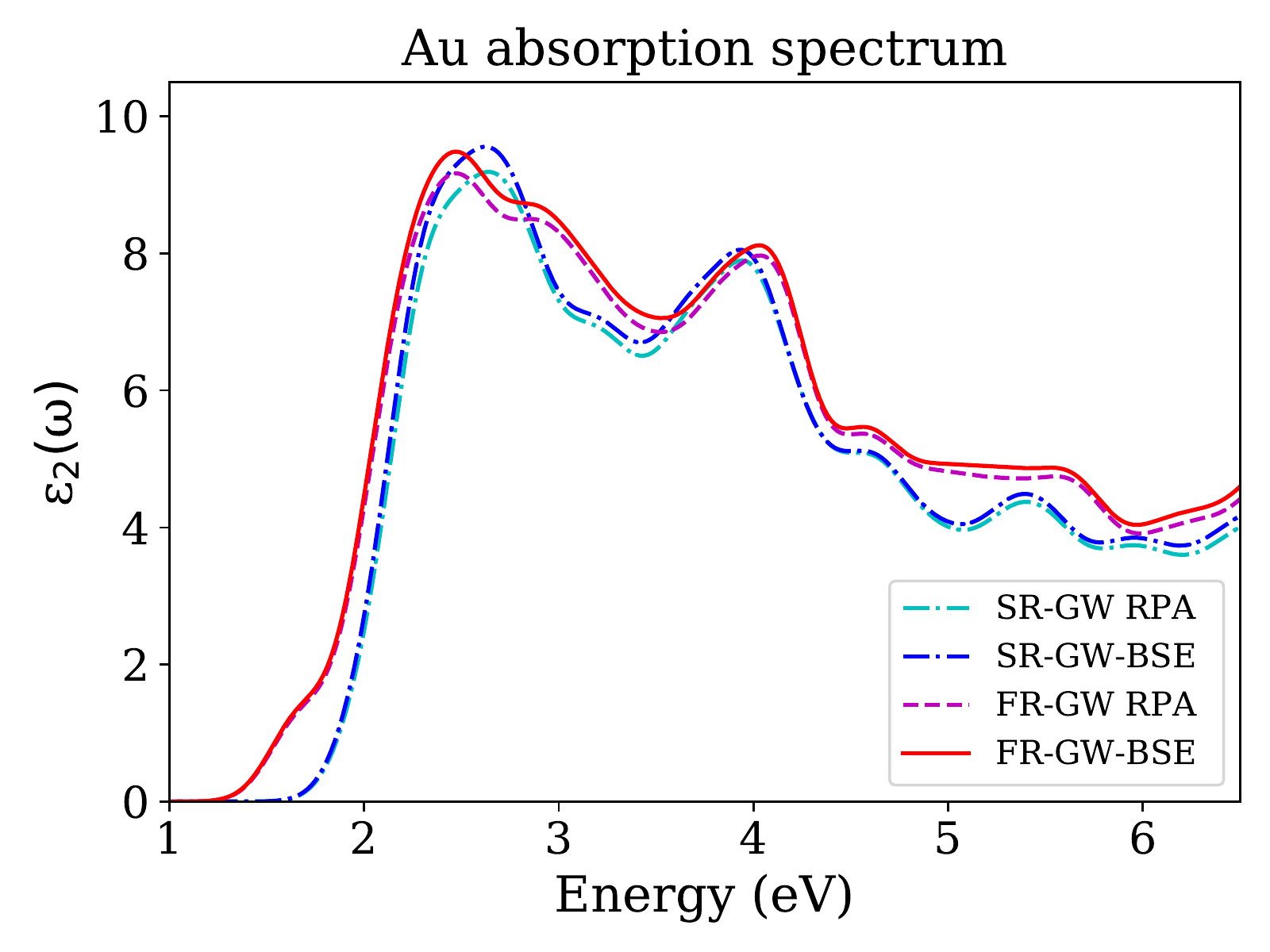}
 \caption{The absorption spectra of Au due to interband transitions, calculated at the GW-BSE level. Spin-orbit is included (neglected)
           in the red (blue) curve.}
  \label{fig:figure_au_absp}
\end{figure}

\subsection{Bi$_2$Se$_3$}
\label{sec:bi2se3}
While the previous test systems confirm the sufficiency of treating SOC as a perturbation, the $GW$+SOC approach for the bandstructure of $\Bi2Se3$ has shown mixed results\cite{yazyev,aguilera}. The large spin-orbit splitting
of the Bi $6p$ electrons inverts the positive and negative parity $p$-like states (from the Bi $6p$ and Se $4p$ orbitals) near the band gap,
creating a nontrivial value of
the $Z_2$ topological index\cite{zhang}. The ``inverted'' band gap is caused by the level-repulsion of the inverted states at the $\Gamma$-point,
mixing the character of the conduction and valence states within a neighborhood of $\Gamma$\cite{yazyev}. The strength of this level repulsion depends on the size of the band gap, which is underestimated in DFT. As a consequence, the bandstructure computed from DFT and $GW$ differ significantly when SOC is included, so the perturbative treatment may be insufficient.



\begin{figure}[p]
  \centering
  \includegraphics[width=0.5\textwidth]{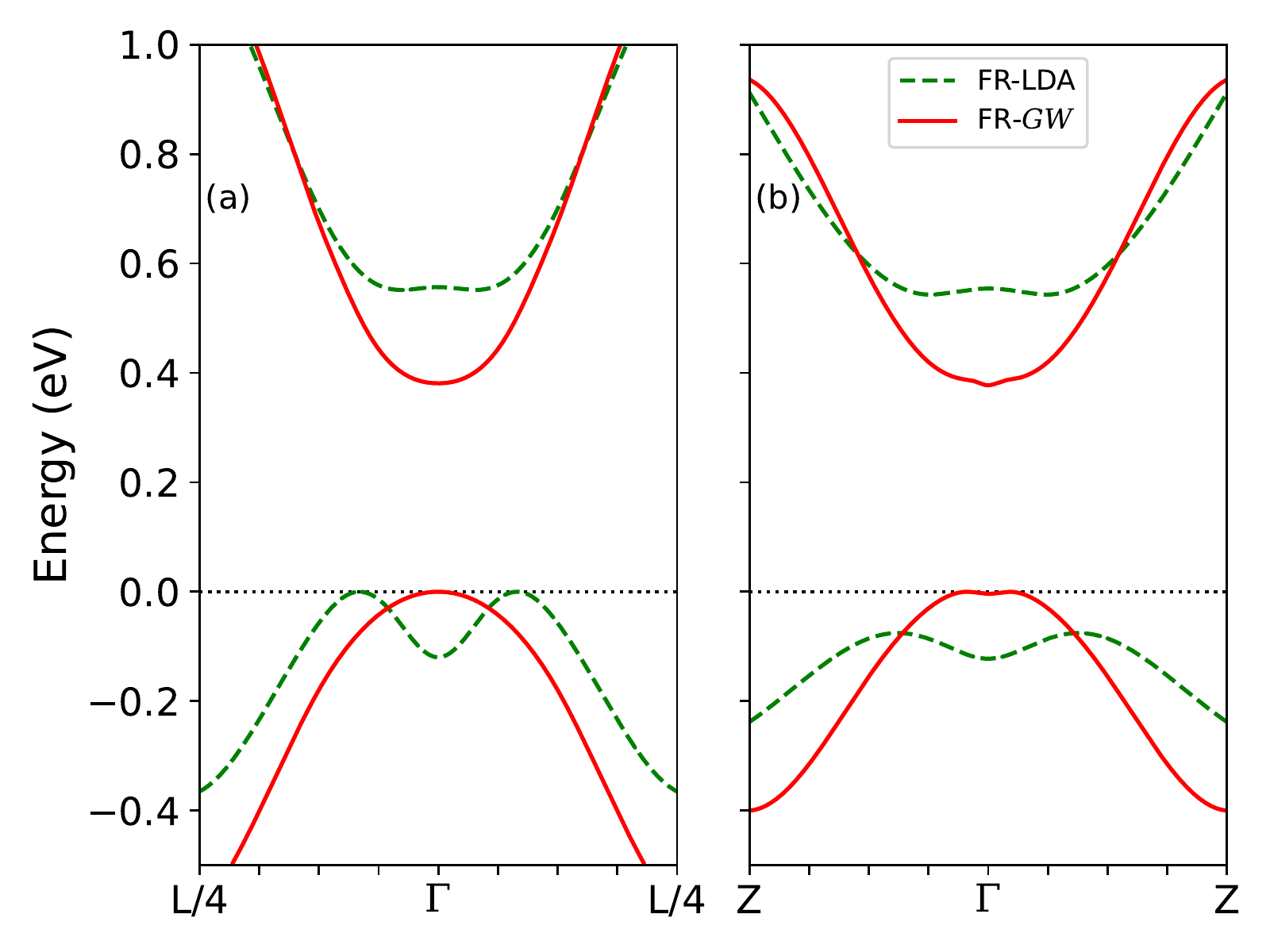}
  \caption{The electronic bandstructure of Bi$_2$Se$_3$ along the (a) $\Gamma$ to $L$  and (b) $\Gamma$ to $Z$  directions, including spin-orbit coupling, but only Hamiltonian matrix elements that are diagonal at the FR-$GW$ level: fully-relativistic (``FR'') LDA and $GW$ in dashed green and solid red lines, respectively.
  }
  \label{fig:bi2se3}
\end{figure}

\begin{figure}[p]
  \centering
  \includegraphics[width=0.5\textwidth]{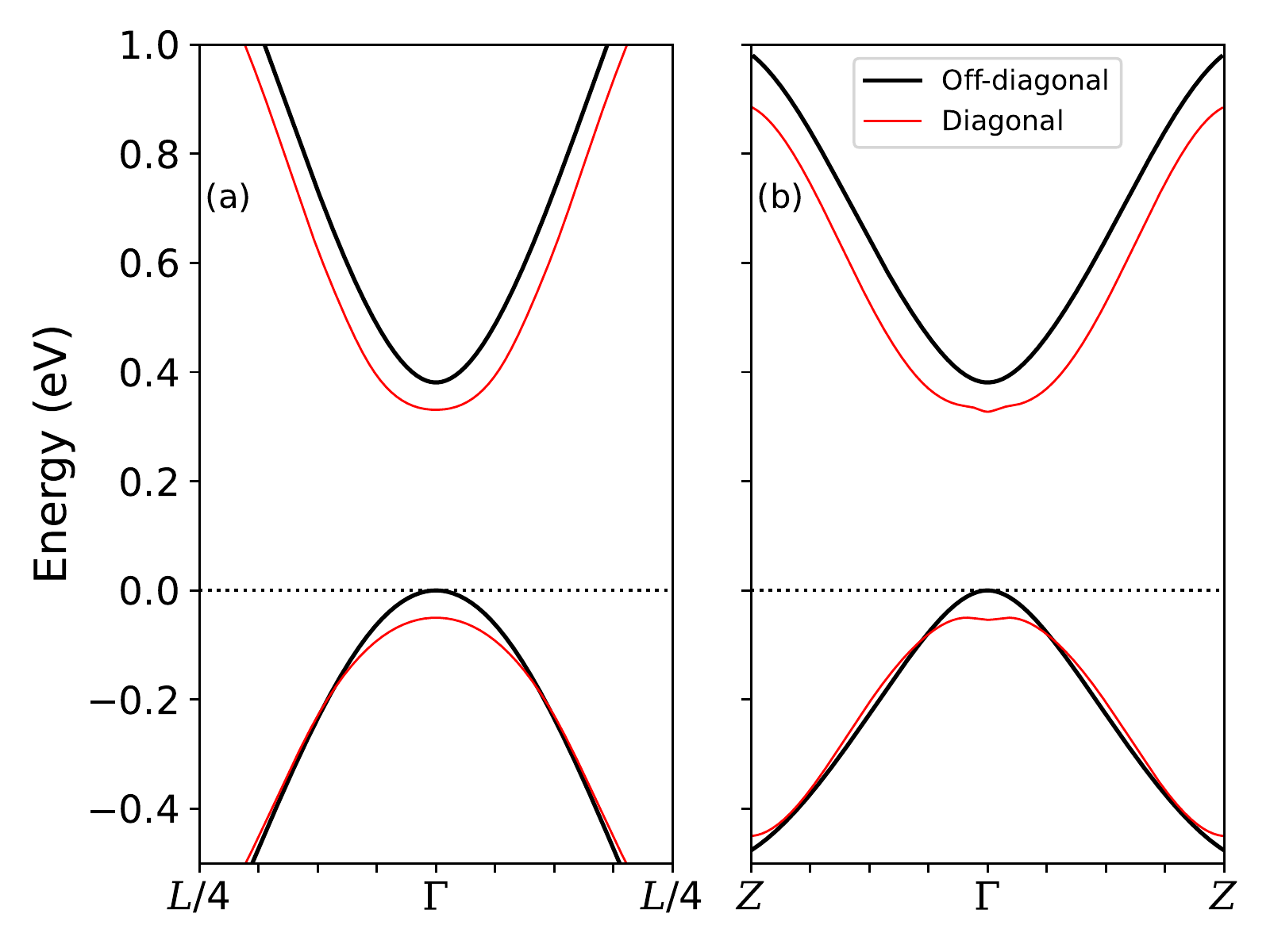}
  \caption{The electronic bandstructure of Bi$_2$Se$_3$ along the (a) $\Gamma$ to $L$ and (b) $\Gamma$ to $Z$ directions, including spin-orbit coupling: the quasiparticle bandstructure computed from FR-$GW$ with off-diagonal entries in the Hamiltonian (solid black), and the quasiparticle bandstructure without off-diagonals (solid, thinner red) and arbitrarily shifted downward by 0.05~{eV} for clarity.
}
  \label{fig:bi2se3_offdiagonal}
\end{figure}

Due to the sensitivity of this system to the DFT functional and the atomic geometry\cite{functional_sensitivity}, in our study of the bulk band gap of $\Bi2Se3$ as computed within FR-$GW$, we use the experimental geometry\cite{madelung}.
For consistency with
the majority of previous calculations in the literature \cite{yazyev,nechaev,aguilera,rohlfing_bi2se3}, we use the LDA functional. We use a Brillouin zone sampling of 8$\times$8$\times$8 for constructing the charge density
as well as the dielectric function. We use a 160~{Ry} cutoff for the planewave basis for the wavefunctions and a 25~{Ry} cutoff for the dielectric function.
The polarizability (``Chi'') summation uses 1000 unoccupied bands, and the Coulomb-hole (``COH'') summation uses 1254 bands. The quasiparticle energies are estimated to be converged within about 30~{meV} \cite{suppmat}.

We obtain the bandstructure in the neighborhood of the $\Gamma$-point by obtaining quasiparticle energies at the $\Gamma$-point and at particular points along the $\Gamma$-to-$L$
and the $\Gamma$-to-$Z$ high-symmetry lines, shown in Fig. \ref{fig:bi2se3}. Ordinarily, bandstructures are determined from the set of quasiparticle energies computed on coarse, regularly spaced k-point grid and a set of overlap coefficients computed for the bands on the coarse grid and DFT-computed bands that densely sample high-symmetry lines in the Brillouin Zone\cite{BerkeleyGW}. This approach does not work well, however, in cases such as $\Bi2Se3$, especially near the $\Gamma$-point where the DFT bandstructure and the quasiparticle bandstructure disagree significantly. Instead, we directly compute quasiparticle energies along the $\Gamma$-to-$L$ direction at $\frac{1}{16}L$, $\frac{1}{8}L$, $\frac{3}{16}L$, and $\frac{1}{4}L$, and along the $\Gamma$-to-$Z$ direction at $\frac{1}{16}Z$, $\frac{1}{8}Z$, $\frac{3}{16}Z$, $\frac{1}{4}Z$, $\frac{1}{2}Z$, and $Z$. The whole $\Gamma$-to-$Z$ line
is represented as it is a much shorter path in the Brillouin zone than the $\Gamma$-to-$L$ line. We then plot spline-interpolated curves as estimates
to the quasiparticle bandstructure. The LDA bandstructure interpolated in a similar fashion shows good agreement with the bandstructure calculated explicitly at each k-point \cite{suppmat}.
The (band-diagonal) FR-$GW$ bandstructure along the $\Gamma$-to-$Z$ line suggests that the band-diagonal approximation is not generally sufficient, as is apparent from
the appearance of small, spurious\cite{nechaev,aguilera,piot} bumps in both the conduction and valence bands in a very narrow region about $\Gamma$ (Fig. \ref{fig:bi2se3}).

We find a direct bulk band gap of 0.38 eV in the band-diagonal approximation, which is in good agreement with values obtained from angle-resolved photo-emission spectroscopy (ARPES)\cite{nechaev} (0.332~{eV})
as well as scanning tunneling spectroscopy (STS)\cite{kim} (0.3~{eV}).
Optical measurements of the gap, however, report a smaller value of 0.2~{eV}\cite{piot} and also confirm a direct band gap at $\Gamma$.

To improve the quasiparticle bandstructure, we investigate the effect of including band-off-diagonals in the calculation of the self-energy matrix elements:
\begin{equation}
\label{eq:offdiag}
E_{n\kp} = \textrm{Eig}\left(\epsilon_{l\kp}\delta_{lm} + \langle l,\kp,\alpha | \Sigma_{\alpha\beta}\left(E_{p\kp}\right) - V^{\textrm{xc}}\delta_{\alpha\beta} | m,\kp,\beta \rangle\right),
\end{equation}
where ``Eig'' denotes the eigenvalues of the matrix constructed from the self-energy in the Kohn-Sham orbital basis, the band $n$ is a member of the set of bands spanned by all choices for the indices $l$ and $m$,
and the energy $E_{p\kp}$ at which the self-energy operator is evaluated is chosen from either the row ($E_{l\kp}$) or column ($E_{m\kp}$), as the difference in eigenvalues from this choice and an explicitly-constructed Hermitian matrix for the self-energy correction,
\begin{equation}
\frac{1}{2}(\langle l,\kp,\alpha | \Sigma_{\alpha\beta}\left(E_{l\kp}\right) - V^{\textrm{xc}}\delta_{\alpha\beta} | m,\kp,\beta \rangle + \langle l,\kp,\alpha | \Sigma_{\alpha\beta}\left(E_{m\kp}\right) - V^{\textrm{xc}}\delta_{\alpha\beta}| m,\kp,\beta \rangle),
\end{equation}
as used for quasiparticle self-consistent GW\cite{qsgw_kotani}, is found to be under 1~{meV}.

We find that the choice of the four valence bands and two conduction bands near the Fermi energy is sufficient to correct the deficiencies in the bandstructure when using the LDA eigenfunctions
as the quasiparticle wavefunctions \cite{suppmat}. The bandstructure computed in this fashion is shown in Fig. \ref{fig:bi2se3_offdiagonal}. The band gap computed at first iteration is 0.33~{eV}, though we rigidly shift the gap to match that of the diagonal approximation, 0.38~{eV}, which is justified in the discussion after Eq. \ref{eq:Eqp_Z}. The necessity of calculating a matrix for the self-energy can be seen by noting that the strength of the level repulsion -- and therefore the character
of the wavefunctions -- depends on the band gap value\cite{yazyev}. When changing the gap, as in a $GW$ calculation, the wavefunctions in the region where the
character is inverted necessarily change along with the extent of the region in the bandstructure with inverted orbital character.
The use of the LDA basis, then, is not adequate for an accurate bandstructure in this region.

In the usual band-diagonal approximation to the self-energy operator two energies are calculated \cite{BerkeleyGW}. First, the self-energy operator is evaluated at the mean-field eigenvalues, giving the first of these energies:
\begin{equation}
\label{eq:Eqp0}
E^0_{n\kp} = \epsilon^{\textrm{DFT}}_{n\kp} + \langle n\kp | \Sigma (\epsilon^{\textrm{DFT}}_{n\kp}) - V^{\xc} | n\kp \rangle.
\end{equation}
$\Sigma$ is evaluated (within default settings in BerkeleyGW) at $\epsilon^{\textrm{DFT}}_{n\kp}$ and $\epsilon^{\textrm{DFT}}_{n\kp} + 1$~{eV}, from which the derivative $\frac{\d \Sigma_{n\kp}}{\d E}$ and the renormalization factor
\begin{equation}
Z_{n\kp} = \frac{1}{1 - \frac{\d \Sigma_{n\kp}}{\d E}}
\end{equation}
are computed. The quasiparticle energy $E^{\textrm{QP}}_{n\kp}$ can then be determined from Newton's Method, and written as
\begin{equation}
\label{eq:Eqp_Z}
E^{\textrm{QP}}_{n\kp} = Z_{n\kp} E^0_{n\kp} + (1 - Z_{n\kp}) \epsilon^{\textrm{DFT}}_{n\kp}.
\end{equation}
However, in band-offdiagonal calculations, the renormalization factors $Z$ cannot be computed in this way and thus the Newton's Method approach to determining the quasiparticle energy cannot be performed. 
The self-energy operator is a function of energy, and a solution to Dyson's equation is found when this input energy is the same as the eigenvalue of the Hamiltonian in Eq. \ref{eq:offdiag}. Initially, the Kohn-Sham energy eigenvalues are used as input for the self-energy, and the basis set is taken to be the Kohn-Sham bands. After diagonalization, a new set of energy eigenvalues and bands (expressed as a linear combination of the Kohn-Sham bands) are used to construct a new Hamiltonian, which is then diagonalized. This process is repeated until the energy eigenvalues do not substantially change from one iteration step to another -- only at that final iteration are the energy eigenvalues the quasiparticle energies.

After a first diagonalization of the Hamiltonian constructed with Kohn-Sham energies and bands, the difference between valence band maximum and conduction band minimum at the $\Gamma$-point is 0.33~{eV}. This energy, though, is analogous to the $E^0_{n\kp}$ energies in Eq. \ref{eq:Eqp0}.
 However, since all off-diagonal terms for the self-energy at the $\Gamma$-point for $\Bi2Se3$ are found to be zero within numerical precision, the self-consistently calculated quasiparticle energies must match exactly at the $\Gamma$-point. We use this fact to rigidly shift the conduction band from the off-diagonal calculation to match the quasiparticle band gap computed when neglecting off-diagonal components, 0.38~{eV}: $E^{\textrm{\textrm{QP}}}_{c\vk} \approx E^{\textrm{off-diag}}_{c\vk} + (E^{\textrm{QP, diag}}_{c,\textbf{k} = \Gamma} - E^{\textrm{off-diag}}_{c,\textbf{k}=\Gamma}) $. This is expected to be acceptable when the off-diagonal matrix elements of the self-energy for the states away from the $\Gamma$-point are sufficiently weakly sensitive to corrections to the Kohn-Sham eigenvalues.

As seen in Fig. \ref{fig:bi2se3_offdiagonal} the conduction and valence bands are now unambiguously parabolic after updating the basis set, so we can readily compute the effective masses. 
We calculate an effective mass of 0.19~{$m_e$} for the holes and 0.14~{$m_e$} for the electrons, averaging over the directions plotted. This compares favorably with the experimentally determined effective masses, from magneto-optics, of 0.14~{$m_e$} for both the electrons and holes \cite{piot}. We note that our determination of effective masses agrees despite the discrepancy in the value of the band gap.

To investigate the sensitivity of the bandgap to the treatment of dynamics in the self-energy operator, we also calculate the band gap at the $\Gamma$-point through the use of the full-frequency treatment of the dielectric function, via the contour deformation method\cite{contour_deformation} and a low rank approximation\cite{low_rank,govoni_galli,mauro}. We used 15 imaginary frequencies, 200 eigenvectors in a reduced basis scheme, corresponding to roughly 10 percent of the full spectrum, a frequency spacing of 0.25 Ry, with frequencies calculated out to 10 Ry. 
We found a slightly larger gap than in the Hybertsen-Louie GPP, with a value of 0.41~{eV}. In conventional semiconductors, redistribution of the weight of the screening from a single frequency typically results in a lower gap; the increase of the gap for $\Bi2Se3$ relative to the GPP result is understood as a consequence of the inverted nature of the bandstructure. The small change in the value indicates that the use of a GPP model for the dynamics in the self-energy is sufficiently accurate for quasiparticle energies.

We can compare our FR-$GW$ calculations of the bandstructure to a $GW$+SOC calculation\cite{yazyev} performed with BerkeleyGW. The band gap in $GW$+SOC is found to be direct at $\Gamma$, with a value of 0.33~{eV}, with parabolic valence and conduction bands. (See Table \ref{tab:bi2se3_comparison} for computational details.) In all FR-$GW$ cases, the quasiparticle bandgap is found to be in a range between 0.38~{eV} to 0.41~{eV}, depending on the treatment of the frequency-dependence of the self-energy operator, and an update to the quasiparticle basis set is required to recover parabolic bands for the valence band maximum and conduction band minimum. Unlike the perturbative $GW$+SOC approach, FR-$GW$ more readily allows for a quasiparticle self-consistent approach to arrive at a quasiparticle basis set in which the dependence on qualitatively inaccurate starting-point mean-field bands 
is removed.


\section{Comparison with other implementations}
\label{sec:comparison}

Other excited-state $GW$ codes have implemented compatibility with spinor wavefunctions, including pseudopotential plane-wave codes WEST\cite{west} and Yambo\cite{yambo}, pseudopotential PAW codes VASP\cite{vasp} and GPAW\cite{gpaw_soc}, and all-electron codes FHI-AIMS\cite{fhiaims}, Questaal\cite{questaal_thesis}, TURBOMOLE\cite{turbomole}, and SPEX\cite{spex,sakuma}.
At present, only Yambo\cite{yambo,yambo_bse_soc} and BerkeleyGW have BSE with spinor wavefunctions implemented. We compare our present results with these other implementations, as a first attempt at benchmarking spinorial $GW$ and $GW$-BSE calculations in the spirit of the $GW$100 set\cite{GW100} and the community effort to examine reproducibility of $G_0W_0$ calculations in solids\cite{reproducibility}. For the standard semiconductors Si, GaP (parameters in Table \ref{tab:parameters}), and GaAs, we find good agreement (in Table \ref{tab:west}) with the computed spin-orbit splitting at the valence band maximum as computed in Ref.\cite{west}, despite that work's use of different pseudopotentials from the SG15 database\cite{sg15} with PBE exchange-correlation functionals\cite{pbe}. The differences between the present calculations and that of WEST  for GaAs are larger than that of Si and GaP due to the considerable underestimation of the direct band gap for GaAs at the SR-$GW$ level computed in Ref. \cite{west} using pseudopotentials from the SG15 database (0.62~{eV}), compared to that from the Pseudo-Dojo database (1.26~{eV}), a discrepancy not present in their results for Si and GaP.  Also apparent in Table \ref{tab:west} is agreement in the shifts of the band gaps upon inclusion of spin-orbit coupling to tens of meV. Results were not found in the literature for FR-$GW$/BSE calculations of these materials as a comparison. Yambo's spin-orbit implementation paper \cite{yambo19} shows results on 2D transition-metal dichalcogenides only, and YAMBO results for GaSb are available only as an unconverged tutorial example\cite{yambo_gasb_tutorial}.

\begin{table}[p]
\begin{ruledtabular}
\caption{
\label{tab:west}
The spin-orbit splitting at the valence band maximum and the change of band gap upon inclusion of spin-orbit coupling for Si and GaAs, in comparison to results computed in the code WEST\cite{west}.}
\begin{tabular}{lccccc}
 & \multicolumn{2}{c}{$\Delta^{\SOC}(\Gamma,v)$ (eV)} &  \multicolumn{2}{c}{$E^{\textrm{FR}}_g - E^{\textrm{SR}}_g$ (eV)} & \multicolumn{1}{c}{$E_{g}$ (eV)} \\\cline{2-3}\cline{4-5}\cline{6-6}
 &  FR-LDA & FR-$GW$ & DFT & $GW$ & FR-$GW$\\
\hline
Si, present & 0.047 & 0.049 & -0.016 & -0.016  & 1.22 \\
Si, Ref. \cite{west} & 0.048  & 0.049  &  -0.016 & -0.017 & 1.36 \\
\hline
GaP, present & 0.089 & 0.086 & -0.027  & -0.024 & 2.57\\
GaP, Ref. \cite{west} & 0.083 & 0.092 & -0.028 & -0.031 & 1.91 \\
\hline
GaAs, present & 0.320 & 0.340 & -0.098  & -0.109 & 1.49\\
GaAs, Ref. \cite{west} & 0.328  & 0.344  & -0.136  &  -0.123 & 0.13\\
\end{tabular}
\end{ruledtabular}
\end{table}

Several results have been reported in the literature for the bulk quasiparticle bandgap for $\Bi2Se3$, from both $GW$+SOC and FR-$GW$ approaches. The $GW$ implementation in Yambo\cite{yambo,yambo19}, a plane-wave pseudopotential excited-state code that computes the polarizability and self-energy with sums over empty states, is most directly comparable to BerkeleyGW, and we find good agreement for our computed FR-$GW$ results for the band gap in the diagonal approximation: 0.36~{eV} from Ref. \cite{yambo_bi2se3}, and 0.38~{eV}, present work. The quasiparticle bandstructure in Ref. \cite{yambo_bi2se3} suggests a direct gap at $\Gamma$, though the resolution is not fine enough to determine if the bands have a parabolic dispersion. The present calculation of the band gap differs only by 20 meV from a prior calculation using BerkeleyGW employing the ``$GW$+SOC'' approach\cite{yazyev}, in which spin-orbit coupling was added perturbatively after evaluating quasiparticle energies that neglected spin.
The bandstructure reported in Ref. \cite{rohlfing_bi2se3} uses pseudopotentials without semicore Bi orbitals, with both Gaussian orbital and plane-wave basis sets in separate calculations. This bandstructure is computed with a non-uniform sampling of the Brillouin Zone for evaluation of self-energy matrix elements, up to 78$\times$78$\times$1 near the zone-center, featuring a direct gap of 0.20 eV and valence band maximum at the $\Gamma$-point with a flattened parabolic shape.
FR-$GW$ FLAPW calculations\cite{nechaev,aguilera} using the SPEX\cite{spex,sakuma} code show a sensitivity to the band gap to calculation parameters at both the DFT and $GW$ levels. Changing the number of local orbitals from 1\cite{nechaev} to 2\cite{aguilera}, $l_{max}$ for $GW$ from 4\cite{nechaev} to 5\cite{aguilera}, planewave cutoff for $GW$ from 3.5 bohr$^{-1}$\cite{nechaev} to 2.9 bohr$^{-1}$\cite{aguilera}, and number of empty states from 300\cite{aguilera} to 500\cite{nechaev} changes the band gap from 0.34~{eV} to 0.20~{eV}. By contrast, a perturbative $GW$+SOC calculation\cite{aguilera} with FLAPW found a vanishing band gap, and the bands appeared to become linear, unexpected for the bulk material.
To conclude, different descriptions of the wavefunctions for $\Bi2Se3$ can give bandgaps of that vary from about 0.2 to 0.35~{eV}, with gaps in this range justified by experiments\cite{nechaev,kim,piot}. Further study regarding self-consistent updates to the quasiparticle wavefunctions within the planewave pseudopotential FR-$GW$ approach can elucidate the features of the bands near the Fermi energy responsible for this sensitivity, and more clarity on the disagreement from ARPES\cite{nechaev} and transmissivity measurements\cite{piot} is needed.
A comparison between these Bi$_2$Se$_3$ calculations is presented in Table \ref{tab:bi2se3_comparison}.

\begingroup
\squeezetable
\begin{table}[p]
\begin{ruledtabular}
\caption{
\label{tab:bi2se3_comparison}
Comparison between present FR-$GW$ and other excited-state calculations from the literature for the bulk band gap for $\Bi2Se3$.
PW = plane wave, PSP = pseudopotential, GPP = generalized plasmon pole, GN = Godby-Needs plasmon pole, CD = contour-deformation.}
\begin{tabular}{lccccccccc}
            & Structure & SOC        & Basis & XC         & Grid,         & Grid,        & No. Empty & Frequency  & Band    \\
            &           & Treatment  & Set   & Functional & Polarization  & Self-energy  & States    & Dependence & Gap (eV)\\
\hline
present & expt. & FR-$GW$ & PW PSP & LDA & 8$\times$8$\times$8 & 8$\times$8$\times$8 & 1254 & GPP & 0.38\\
Ref. \cite{yazyev} & expt. & $GW$+SOC & PW PSP & LDA & 6$\times$6$\times$6 & 6$\times$6$\times$6 & $\sim$500 & GPP & 0.36\\
Ref. \cite{yambo_bi2se3} & relaxed & FR-$GW$ & PW PSP & PBE & 6$\times$6$\times$6 & 6$\times$6$\times$6 & 3000 & GN & 0.36\\
Ref. \cite{nechaev} & expt. & FR-$GW$ & FLAPW & LDA & 4$\times$4$\times$4 & 4$\times$4$\times$4 & 300 & CD & 0.34\\
Ref. \cite{aguilera} & expt. & FR-$GW$ & FLAPW & LDA & 4$\times$4$\times$4 & 4$\times$4$\times$4 & 500 & CD & 0.20\\
Ref. \cite{aguilera} & expt. & $GW$+SOC & FLAPW & LDA & 4$\times$4$\times$4 & 4$\times$4$\times$4 & 500 & CD & 0.01\\
Ref. \cite{rohlfing_bi2se3} & expt. & FR-$GW$ & PW/Gaussian PSP & LDA & 8$\times$8$\times$8 & Non-uniform\footnote{10$\times$10$\times$1 to 78$\times$78$\times$1} & 234 & GN & 0.20\\
\end{tabular}
\end{ruledtabular}
\end{table}
\endgroup

\section{Performance}
\label{sec:performance}

We give a comparison in the performance of BerkeleyGW for the representative case of GaAs with and without spinor wavefunctions. We see in Table \ref{tab:performance} that the calculation of the wavefunctions for the self-energy matrix elements (``DFT Coarse'') takes four times longer, in accordance with expectation from having to double the number of bands and double the size of each band, for the spin-up and spin-down components. The calculation of the wavefunctions for the basis of the BSE Hamiltonian (``DFT Fine'') is more rapid, since the bottleneck in generating these wavefunctions is the number of k-points and not the number of bands. Calculation of the dielectric matrix (``\texttt{Epsilon}'') sees an increase in cost of only 2.5, far less than the increase in cost of generating the matrix elements alone, because the matrix inversion step is a significant fraction of runtime, and it is unaffected by spinors since the size of the dielectric matrix is the same when including or disregarding spin. The calculation of quasiparticle energies (``\texttt{Sigma}''), however, is closer to the expected increase in cost, at a factor of 4.1. The costs of constructing the BSE kernel (``\texttt{Kernel}'') and solving the eigenvectors and eigenvalues (``\texttt{Absorption}'') have the largest increases, at 6.4 and 15.0, respectively. The \texttt{Kernel} code requires the calculation of three sets of matrix elements, an increase in cost partially offset by time spent on the better-scaling routines such as I/O. We discuss the Absorption code performance in more detail below.

The \texttt{Absorption} code has four main routines: I/O, interpolation of the quasiparticle energies, interpolation of the kernel matrix elements, and diagonalization. We see the performance of each when disregarding spin and when using spinor wavefunctions in Table \ref{tab:absorption_performance}. The I/O necessarily has an increase in cost of a factor of 4, from the increase in the size of the wavefunction files. Similarly, the interpolation of the quasiparticle energies takes nearly 4 times longer, for the same reason. The interpolation of the kernel matrix elements increases cost by a factor of 10.1, less than an estimated increase of 16, since the interpolation coefficients have been calculated in the previous step, and the multiplication with the kernel matrix elements is performed as an optimized matrix-matrix multiplication with the Level 3 BLAS call ZGEMM\cite{BerkeleyGW}. The diagonalization sees an increased cost by a factor of 56.8, close to the expected factor of 64.

\begin{table*}[p]
\begin{ruledtabular}
\caption{\label{tab:performance}
Comparison of performance of BerkeleyGW on GaAs, when disregarding spin and when using spinor wavefunctions.}
\begin{tabular}{ccccc}
Step &  No. CPUs & CPU Hours (no spin) & CPU Hours (spinor) & Ratio\\
\hline
DFT Coarse & 1024 & 162 & 650 & 4.0\\
DFT Fine & 1728 & 173 & 490 & 2.8\\
\texttt{Epsilon} & 864 & 864 & 2160 & 2.5\\
\texttt{Sigma} & 864 & 2760 & 11232 & 4.1\\
\texttt{Kernel} & 600 & 560 & 3600 & 6.4\\
\texttt{Absorption} & 600 & 48 & 720 & 15.0\\

\end{tabular}
\end{ruledtabular}

\end{table*}

\begin{table*}[p]
\begin{ruledtabular}
\caption{\label{tab:absorption_performance}
Comparison of performance of \texttt{Absorption} executable in BerkeleyGW when disregarding spin and when using spinor wavefunctions, as seen in calculations of GaAs.}
\begin{tabular}{cccc}
Step &  Wall time, no spin (s) & Wall time, spinor (s) & Ratio\\
\hline
I/O & 138 & 560  & 4.0\\
Interp. WFN & 57 & 240 & 4.2\\
Interp. Kernel & 27 & 274 & 10.1\\
Diag.  & 53 & 3013 & 56.8\\
\end{tabular}
\end{ruledtabular}
\end{table*}

\section{Conclusion}
\label{sec:conclusion}

Our implementation of spinor GW/BSE in the BerkeleyGW excited-state software enables computation of the quasiparticle energies and absorption spectra for materials with large SOC. The use of DFT one-particle wavefunctions with two spinor components necessarily increases the cost of calculation, found in practice to be at best about three times more expensive than when neglecting SOC, and with the calculations necessary for calculating the optical absorption within the $GW$-BSE being much more expensive due to the increase in basis set size. The careful use of symmetry however can significantly reduce the cost in some systems.

We demonstrated our implementation on the test systems Si, Ge, GaAs, GaSb, CdSe, and Au, which were readily calculated in the band-diagonal, one-shot $G_0W_0$ method. The band gaps, spin-orbit splittings, and energy eigenvalues were shown to be highly accurate across this range of different spin-orbit coupling strengths. 
The band gaps were also shown to be well-approximated when introducing SOC as a perturbation to the valence band maxima computed while neglecting spin.
The topological insulator material $\Bi2Se3$, however, needed some correction to the LDA basis for the quasiparticle states. While a fairly accurate band gap of 0.38~{eV}
was computed within band-diagonal $G_0W_0$, the bandstructure shows small but unphysical features in a small neighborhood about the $\Gamma$-point. We demonstrated that correcting the LDA basis states by diagonalizing the $G_0W_0$ Hamiltonian was able to remove this unphysical feature, and provide effective masses in good agreement with experiment.

We additionally performed fully-relativistic Bethe Salpeter Equation calculations
of the absorption spectra for GaAs and GaSb. We show that the absorption spectrum
for GaAs is similar within both the SR-$GW$-BSE and FR-$GW$-BSE. For GaSb we are able to resolve the spin-orbit split $E_1$ and $E_1+\Delta$ peaks, with their placement within tens of meV of experiment.

The perturbative treatment of spin-orbit coupling for electronic structure, $GW$+SOC, shows high agreement with the more costly non-perturbative FR-$GW$ approach for many test systems. Such systems, even with nominally strong spin-orbit coupling as in GaSb and Au, have fully-relativistic DFT bandstructures that have high qualitative agreement with that from FR-$GW$. However, for materials such as $\Bi2Se3$ that possess both a narrow bandgap and strong spin-orbit coupling, the significant qualitative differences between the fully-relativistic DFT bandstructures and FR-$GW$ motivate the use of the FR-$GW$ approach. $GW$+SOC approaches for $\Bi2Se3$ have shown conflicting qualitative descriptions of the bulk bandgap\cite{yazyev,aguilera}, while the FR-$GW$ approaches\cite{yambo_bi2se3,nechaev,aguilera,rohlfing_bi2se3} have been consistent, within about 0.1~{eV}. Further, the use of FR-$GW$ allows for updating the quasiparticle wavefunctions, which then gives good quantitative agreement with the experimentally measured effective masses for electrons (0.14~{$m_e$, experiment\cite{piot} and computed} and holes (0.14~{$m_e$}, experiment\cite{piot} and 0.19~{$m_e$}, computed) for $\Bi2Se3$. The use of FR-$GW$-BSE for the test systems of GaAs and GaSb considered presently gives no significant advantage\cite{cohen_epm_soc} over the perturbative approach\cite{felipe_diana}, and results on monolayer transition metal dichalcogenides in the literature also show agreement to a few 10~{meV} between the non-perturbative and perturbative inclusion of SOC in the $GW$-BSE excitonic binding energies \cite{yambo_bse_soc}. However, it is reasonable to think that in materials where SOC gives a qualitative difference in bandstructure,
like $\Bi2Se3$, there may be stronger effects in BSE not captured by a perturbative treatment.

The availability of spinor $GW$/BSE calculations in BerkeleyGW opens the way to increased use of fully relativistic quasiparticle and excitonic absorption calculations in the electronic structure community, enabling more accurate and detailed exploration of topological materials which have garnered great recent research interest, as well as in thermoelectric and photovoltaic materials. BerkeleyGW has particular strengths for large and reduced-dimensional systems, such as a defect in a 2D topological material\cite{chrobak2020fe}. Further developments include the use of magnetic group symmetries to facilitate the computation of non-collinear magnetic systems without the requirement of large supercells\cite{yambo19}, and the calculation of non-collinear spin-susceptibilities\cite{electron-magnon}, as well as more benchmarking FR-$GW$-BSE for materials with large spin-orbit coupling and large exciton binding energies. We find good agreement with other existing $GW$ implementations, and believe that further detailed comparison can help to improve implementations of this methodology and ensure accuracy.

This implementation of spinor $GW$/BSE in BerkeleyGW was publicly released in BGW version 3.0, and a tutorial example for performing $GW$/BSE calculations is available at URL \texttt{https://workshop.berkeleygw.org/tutorial-workshop-resources/about}. 

\section{Acknowledgments}
\label{sec:acknowledgments}
The authors would like to thank Felipe H. da Jornada, Derek Vigil-Fowler,
Georgy Samsonidze, Gabriel Antonius, Tonatiuh Rangel, and Jeffrey B. Neaton for productive discussions. The authors would also like to thank Jamal Mustafa for sharing the private Python repository for \texttt{cms-py}, scripts for computing quasiparticle Fermi energies for metals, used for the quasiparticle bandstructure for Au.
This work was supported by National Science Foundation Grant No. DMR10-1006184 and by the Office of Science, Office of Basic Energy Sciences, U.S. Department of Energy: by the Director of the Materials Sciences and Engineering Division under Contract No. DE-AC02-05CH11231, 
within the Nanomachines Program (KC1203) and within the Theory of Materials Program (KC2301), as well as from the CTC and CPIMS Programs, under Award No. DE-SC0019053.
Computational resources have been provided by the DOE at Lawrence Berkeley National Laboratory's NERSC facility.


%

\end{document}